\documentclass[twocolumn,showpacs,preprintnumbers,nofootinbib,prd,
superscriptaddress,10pt]{revtex4-1}

\usepackage{amsmath,amssymb}
\usepackage[normalem]{ulem}
\usepackage{textcomp}
\usepackage{hyperref}
\usepackage{bm}
\usepackage{graphicx}
\usepackage{psfrag}
\usepackage[usenames,dvipsnames]{xcolor}
\usepackage[utf8]{inputenc}
\usepackage{multirow}
\usepackage{tabularx}

\graphicspath{{./figures/}}
\allowdisplaybreaks[4]

\newcommand{\sub}[1]{_{\text{#1}}}
\newcommand{\super}[1]{^{\text{#1}}}

\newcommand{\ord}[1]{\mathcal{O} \left( #1 \right)}
\newcommand{\HH}{\mathcal{H}}

\newcolumntype{C}[1]{>{\centering\arraybackslash}p{#1}}

\begin{document}

\title{The peculiar acceleration of stellar-origin black hole binaries:\\ measurement and biases with LISA}

\date{\today}

\author{Nicola Tamanini}
\affiliation{Max-Planck-Institut für Gravitationsphysik, Albert-Einstein-Institut, Am Mühlenberg 1, 14476 Potsdam-Golm, Germany}

\author{Antoine Klein}
\affiliation{CNRS, UMR 7095, Institut d'Astrophysique de Paris, 98 bis Bd Arago, 75014 Paris, France}
\affiliation{Institute for Gravitational Wave Astronomy and School of Physics and Astronomy, University of Birmingham, Birmingham B15 2TT, United Kingdom}

\author{Camille Bonvin}
\affiliation{D\'{e}partment de Physique Th\'{e}orique and Center for Astroparticle Physics, Universit\'{e} de Gen\`{e}ve, 24 quai Ernest-Ansermet, CH-1211 Gen\`{e}ve 4, Switzerland}

\author{Enrico Barausse}
\affiliation{CNRS, UMR 7095, Institut d'Astrophysique de Paris, 98 bis Bd Arago, 75014 Paris, France}
\affiliation{SISSA, Via Bonomea 265, 34136 Trieste, Italy and INFN Sezione di Trieste}
\affiliation{IFPU - Institute for Fundamental Physics of the Universe, Via Beirut 2, 34014 Trieste, Italy}

\author{Chiara Caprini}
\affiliation{Laboratoire Astroparticule et Cosmologie, CNRS UMR 7164, Université Paris-Diderot, 10 rue Alice Domon et Léonie Duquet, 75013 Paris, France}

\begin{abstract}
We investigate the ability of the Laser Interferometer Space Antenna (LISA) to measure the center of mass acceleration of stellar-origin black hole binaries emitting gravitational waves.
Our analysis is based on the idea that the acceleration of the center of mass induces a time variation in the redshift of the gravitational wave, which in turn modifies its waveform.
We confirm that while the cosmological acceleration is too small to leave a detectable imprint on the gravitational waveforms observable by LISA, larger peculiar accelerations may be measurable for sufficiently long lived sources.
We focus on stellar mass black hole binaries, which will be detectable at low frequencies by LISA and near coalescence by ground based detectors.
These sources may have large peculiar accelerations, for instance, if they form in nuclear star clusters or in AGN accretion disks.
If that is the case, we find that in an astrophysical population calibrated to the LIGO-Virgo observed merger rate, LISA will be able to measure the peculiar acceleration of a small but significant fraction of the events if the mission lifetime is extended beyond the nominal duration of 4 years.
In this scenario LISA will be able to assess whether black hole binaries form close to galactic centers, particularly in AGN disks, and will thus help discriminate between different formation mechanisms. 
Although for a nominal 4 years LISA mission the peculiar acceleration effect cannot be measured, a consistent fraction of events may be biased by strong peculiar accelerations which, if present, may imprint large systematic errors on some waveform parameters.
In particular, estimates of the luminosity distance could be strongly biased and consequently induce large systematic errors on LISA measurements of the Hubble constant with stellar mass black hole binaries. 
\end{abstract}


\maketitle

\section{Introduction}

The long awaited detection~\cite{LIGO1,boxingday,LIGOJan2017,LIGOJun2017} of gravitational waves (GWs) by the LIGO interferometer, followed by Virgo~\cite{3detBH,LIGONS}, opened a new era in the history of astronomy. Indeed, not only did these observations provide direct evidence for the existence of GWs (whose existence was previously assessed only indirectly by timing binary pulsar systems~\cite{Taylor:1982zz}) and for the validity of General Relativity (GR) in the highly relativistic strong field limit~\cite{LIGOtestsGR}, but they also shed light on the physics of compact objects (black holes -- BHs -- and neutron stars -- NS) \cite{LIGOScientific:2018mvr,Barack:2018yly}.
Indeed, the coincident detection (to within 1.7 s) of the GW signal GW170817 and the gamma ray burst (GRB) 170817A has substantiated the long suspected connection between NS mergers and short GRBs~\cite{Monitor:2017mdv}. Moreover, these joint observations have allowed constraints to be placed on the cornerstones of our understanding of gravity (Lorentz symmetry and the equivalence principle)~\cite{Monitor:2017mdv}, as well as on whole classes of gravitational theories modifying and/or extending GR (see e.g.~\cite{dima,tessa,Sakstein:2017xjx}). GW observations have also started testing the very existence of BHs, down almost to the scale of the event horizon~\cite{Cardoso:2016rao,Barausse:2014tra}, where exotic non-GR physics is still possible, though tightly constrained~\cite{Barausse:2018vdb,Johnson-McDaniel:2018uvs,Westerweck:2017hus}. Furthermore, as more and more BH detections accumulate, GW interferometers will reconstruct the mass function of these objects with increasing precision. This, together with future measurements of the eccentricity and spin of these binaries, will lead to better understanding of their astrophysical origin (see e.g.~\cite{Rodriguez:2016vmx,Nishizawa:2016jji}).

Given this tumultuous succession of discoveries, it is hardly surprising that the European Space Agency (ESA) approved in June 2017 the Laser Interferometer Space Antenna (LISA) mission~\cite{Audley:2017drz}, whose launch is nominally scheduled for 2034 and which will target the mHz band of the GW spectrum. This band of frequencies is inaccessible from the ground, due to the seismic noise affecting the operations of terrestrial interferometers, and contains an impressive wealth of astrophysical GW sources. These include the merger of massive ($\sim 10^4$--$10^7 M_\odot$) BH binaries~\cite{Klein:2015hvg,Tamanini:2016zlh}; Galactic and extra-Galactic white dwarf binaries~\cite{Breivik:2017jip,Korol:2017qcx,Korol:2018wep}; extreme mass ratio inspirals consisting of a stellar-mass BH and a massive BH~\cite{Gair:2017ynp,Babak:2017tow}; and also the stellar-mass BH binaries that are targeted at higher frequencies by LIGO/Virgo~\cite{Sesana:2016ljz}. Indeed, those binaries may be observable by LISA in the tens or even hundreds, especially if their mass is sufficiently large, for several years. These sources will then leave the LISA band before resurfacing in the LIGO/Virgo band, where they will merge~\cite{Sesana:2016ljz}.

While the signal-to-noise ratio (SNR) of these binaries in the LISA band will be at most of the order of a few tens or less (with most sources being indeed unresolvable individually)~\cite{Sesana:2016ljz}, these sources will accrue several thousands inspiral cycles at the high-frequency end of the LISA sensitivity curve. This will potentially result in accurate determinations of the source parameters, such as BH masses and spins, eccentricity, sky position, etc~\cite{DelPozzo:2017kme,Nishizawa:2016jji}. This will also make these sources excellent probes of putative tiny deviations away from GR (orders of magnitude weaker than those detectable with ground interferometers)~\cite{Barausse:2016eii,Vitale:2016rfr} or even environmental effects from the interaction of the binary with the surrounding matter~\cite{Barausse:2014tra,Barausse:2014pra}. 

Among these environmental effects, an especially attractive one is provided by the peculiar acceleration of the binary's center of mass (CoM) with respect to the observer. The peculiar velocity of the binary causes a shift in the signal's frequency, via the well-known Doppler effect. If the velocity is constant during the GW measurement, the Doppler frequency shift can be simply reabsorbed into a change of the (redshifted) chirp mass of the binary and into a shift of the luminosity distance. However, as shown in~\cite{bonvin-2017},  if the velocity varies with time, i.e.~if the binary's center of mass  is accelerating with respect to the observer,  the gravitational waveform changes in a (potentially) measurable fashion. This effect has a characteristic time (or frequency) dependence, and its amplitude is proportional to the magnitude of the acceleration component along the line of sight. The center of mass acceleration depends crucially on the binary's origin: formation in dense nuclear star clusters \cite{OLeary:2008myb,Rodriguez:2016avt,Antonini:2016gqe} or via accretion disks in active galactic nuclei (AGNs) \cite{Bellovary:2015ifg,Bartos:2016dgn,Stone:2016wzz,McKernan:2017umu,Yang:2019okq} provides on average larger accelerations than, for instance, field formation \cite{Inayoshi:2017hgw}. 

In this paper, we investigate the conditions under which the acceleration can be measured by LISA, and its impact on the estimation of the binary parameters. The binary acceleration has two components: one of cosmological origin, due to the expansion of the Universe during the observation of the binary \cite{seto-2001,Nishizawa:2011eq}; and the other of astrophysical origin, i.e.~the peculiar acceleration caused by surrounding matter~\cite{bonvin-2017,Chamberlain:2018snj}.
Our results confirm that the first component is not detectable with LISA \cite{bonvin-2017}, and we thus focus only on the second one\footnote{Note however that a detector with higher sensitivity, like the proposed DECIGO/BBO, has the potential to measure the effect of cosmological expansion~\cite{seto-2001,Nishizawa:2011eq}.}. We assume a constant peculiar acceleration on the timescale of the GW measurement, which is a good assumption for the astrophysical systems considered in this work. 
The magnitude and direction of the acceleration are kept as free parameters.
By adopting the official LISA configuration of \cite{Audley:2017drz}
and astrophysical binary BH populations calibrated to the LIGO/Virgo detection rate, we investigate via a Fisher matrix approach the precision with which the peculiar acceleration can be inferred from the data, improving on the estimates of \cite{bonvin-2017,Inayoshi:2017hgw}. We consider two values for the LISA mission duration: 4 years, the nominal duration, and 10 years, a very plausible extension \cite{Audley:2017drz}.
We assess the improvement in the estimate of the acceleration due to the joint detection of the same event by ground based interferometers. Overall, our results confirm that LISA will be able to observe at least a few events with high enough peculiar acceleration, e.g.~if the latter are produced in dense environments such as nuclear star clusters or AGN disk, and if the mission will last for 10 years.
The detection of such peculiar acceleration by LISA would be extremely valuable, since it may provide information on the binary's formation mechanism, especially if the host galaxy can be identified electromagnetically with sufficient confidence (to identify for example the presence of an AGN or a nuclear star cluster). Note that the sky position of stellar mass BH binaries may be identified to within 10 deg$^2$ or better~\cite{DelPozzo:2017kme}.

In addition, we also investigate the bias that the binary's acceleration would generate on the source parameters, if its effect is not properly accounted for in the  GW waveform. Our analysis, which is based on a Fisher matrix approach and which is valid in the small-bias limit~\cite{Cutler-2007,Kitching:2008eq,Camera:2014sba,Cardona:2016qxn}, extends and completes the investigation performed in~\cite{bonvin-2017}, where the bias was calculated using simplified Monte Carlo simulations. 
We confirm that the bias is typically small if peculiar accelerations are below $\sim10^{-7}$ m/s$^2$, but strong biases (larger than the inferred 1$\sigma$ errors) may be present for binaries with larger peculiar accelerations, particularly those forming in AGN disks and nuclear star clusters, even for a 4 year LISA mission. 
Remarkably, the waveform parameter most strongly biased is the luminosity distance of the source.
This suggests that strong peculiar accelerations can induce a large systematic error on measurements of the Hubble constant with stellar mass BHBs observed by LISA (see e.g.~\cite{DelPozzo:2017kme,Kyutoku:2016zxn}).

The plan of the paper is the following.
In the next section we derive the impact of the peculiar acceleration on the observed waveform associated with stellar mass BHBs.
In section~\ref{sec:Fisher} we describe the procedure to implement the acceleration effect in the source catalogues and in the Fisher matrix code.
In section \ref{sec:results} we present our results and we conclude in section \ref{sec:conclusions}.

\section{The effect of the centre of mass acceleration on the gravitational waveform}
\label{sec:theory}

In this section we show how the acceleration of the center of mass of a binary system emitting GWs affects the waveform signal detected by LISA. We follow the derivation presented in~\cite{bonvin-2017}.
Note that, with respect to~\cite{bonvin-2017}, we concentrate only on the redshift perturbation due to the homogeneous expansion of the universe and the peculiar velocity, which provide the dominant effects on the waveform.
We therefore neglect the cosmological contributions due to the Bardeen potentials and the integrated Sachs Wolfe effect. Furthermore, we correct a sign error in the wave phase, c.f.~Eq.~\eqref{eq:pert_phase_FS}, and add a new contribution in the wave amplitude, c.f.~Eq.~\eqref{eq:amp_corrected_FS}.
In the following we provide the main points leading to the perturbed waveform in Fourier space, while the full derivation is reported in the Appendix.

We start by considering the GW signal in the source frame, which is given by~\cite{Blanchet-LR,Maggiore:1900zz}
\begin{align}
 h_{\rm src}(t_{\rm src}) &= \frac{2 G M \nu y^2}{d_C c^2} \sum_{n \geq 0} A_n e^{-i n \phi_{\rm src}} + c.c., \label{eq:TDWF} \\
  2 \pi f_{\rm src} &= \frac{d \phi_{\rm src}}{dt_{\rm src}} = \frac{c^3}{G M} y^3 \,,
\end{align}
where $M$ is the intrinsic total mass of the system, $\nu = m_1 m_2 / M^2$ is the symmetric mass ratio, $d_C$ is the comoving distance from the source to the detector, $n$ is a harmonic number, $A_n$ is the corresponding wave amplitude, $\phi_{\rm src}$ is the orbital phase (at the source), and $y = (G M \omega_{\rm src} / c^3)^{1/3}$ is a post-Newtonian parameter used to describe the system, with $\omega_{\rm src} = 2 \pi f_{\rm src}$ where $f\sub{src}$ is the orbital frequency of the system in the source frame.
An observer at cosmological distances from the source would measure this signal redshifted by both the background expansion of the Universe and the presence of matter structures between the source and the observer, which modify the frequency at observer as follows
\begin{align}
\label{frequency}
f_{\rm obs}=\frac{f_{\rm src}}{1+z}\ ,
\end{align}
where $z$ is the redshift factor, which is a function of time. The dominant contributions to the redshift are due to the cosmological expansion of the Universe and to the Doppler effect (see e.g.~\cite{peacock:1999})
\begin{equation}
	1+z = \frac{a(t_{\rm obs})}{a(t_{\rm src})} \left[ 1 + \frac{v_{\rm src}^\shortparallel(t_{\rm src})}{c} - \frac{v_{\rm obs}^\shortparallel(t_{\rm obs})}{c} \right] \,,
\end{equation}
where $a(t)$ is the scale factor, $t_{\rm src}$ and $t_{\rm obs}$ denote the local time of the source and of the observer, and $v_{\rm src}^\shortparallel = \mathbf{n} \cdot \mathbf{v}_{\rm src}$ and $v_{\rm obs}^\shortparallel = \mathbf{n} \cdot \mathbf{v}_{\rm obs}$ are their respective peculiar velocity along the line of sight $\mathbf{n}$ (here $\mathbf{n}$ points from the observer to the source). As shown in~\cite{bonvin-2017} the gravitational redshift and integrated Sachs-Wolfe contributions can be safely neglected. 
Moreover given that the phase does not change from the source to the observer, the time intervals are related as follows
\begin{align}
\frac{dt_{\rm obs}}{dt_{\rm src}}=\frac{dt_{\rm obs}}{d\phi_{\rm obs}}\frac{d\phi_{\rm obs}}{d\phi_{\rm src}}\frac{d\phi_{\rm src}}{d t_{\rm src}}=\frac{f_{\rm src}}{f_{\rm obs}}=1+z\, .
\end{align}

\subsection{Constant redshift}

If $z$ is assumed to be a constant, i.e.~if both the expansion of the Universe and the peculiar velocity of the source are constant during the GW  observation\footnote{Note that the peculiar velocity of the observer may vary significantly during the time of observation, but this is automatically accounted for in GW analyses since the intrinsic motion of the detector is known with great precision.}, then the GW signal measured by the observer becomes simply
\begin{align}
 h_{\rm obs}(t_{\rm obs}) &= \frac{2 G M_z \nu y^2}{d_L c^2} \sum_{n \geq 0} A_n e^{-i n \phi_{\rm obs}} + c.c., \\
 2 \pi f_{\rm obs} &= \frac{d \phi_{\rm obs}}{dt_{\rm obs}} = \frac{c^3}{G M_z} y^3 \,,
\end{align}
where $M_z = (1+z) M$ is the \emph{redshifted total mass} and $d_L = (1+z) d_C$ is the \emph{luminosity distance}.
The post-Newtonian parameter is unaffected because
\begin{equation}\label{eq:y3const}
	y^3 = \frac{G M}{c^3} \omega_{\rm src} = \frac{G M_z}{c^3} \omega_{\rm obs} \,.
\end{equation}
At leading post-Newtonian order, the frequency evolution equation in the source frame is given by
\begin{align}
 \frac{d\omega_{\rm src}}{dt_{\rm src}} &= \frac{96 \nu c^6 y^{11}}{5 G^2 M^2} \,,
 \label{eq:f_evol_src}
\end{align}
and in the detector frame the same equation can be expressed as
\begin{align}
 \frac{d\omega_{\rm obs}}{dt_{\rm obs}} &=\frac{96 \nu c^6 y^{11}}{5 G^2 M_z^2}. \label{eq:omegadot}
\end{align}
We can use this equation with the relation between the observed frequency $\omega_{\rm obs}$ and the
PN parameter $y$ to compute the time-frequency relation
\begin{align}
t_{\rm obs} - t_c &= \int_{t_c}^{t_{\rm obs}} dt_{\rm obs}' = \int_{y_c}^{y} \left(\frac{dy'}{dt_{\rm obs}'}\right)^{-1} dy'\nonumber\\
& = - \frac{5 G M_z y^{-8}}{256 \nu c^3} \,,
\label{eq:timefreq}
\end{align}
where $t_c$ denotes the coalescence time in the observer's frame and we have used that the frequency diverges at $t_c$ so that $y(t_c)\rightarrow\infty$.
Similarly, we can compute the observed orbital phase
\begin{align}
\phi_{\rm obs} - \phi_c &= \int_{t_c}^{t_{\rm obs}}\! \omega_{\rm obs} dt_{\rm obs}' = \int_{y_c}^{y} \frac{c^3 (y')^3}{G M_z} \left(\frac{dy'}{dt_{\rm obs}'}\right)^{-1}dy'  \nonumber\\
 &= - \frac{y^{-5}}{32 \nu} \,,\label{eq:orbphase}
\end{align}
where $\phi_c=\phi(t_c)$ denotes the phase at coalescence.
We can then compute the Fourier domain version of this waveform using the stationary phase approximation (see e.g.~\cite{Cutler-1994,Lang-2006,Lang-2006-E1,Lang-2006-E2})
\begin{align}
 \tilde{h}_{\rm obs}(f_{\rm obs}) &= \label{eq:SPA}\\
 & \hspace{-.6cm} \frac{2 G M_z \nu}{d_L c^2} \sum_{n \geq 1} \sqrt{\frac{2 \pi}{n|\ddot{\phi}_{\rm obs}|}} A_n y_n^2 e^{i (2 \pi f_{\rm obs} t_n - n \phi_{\rm obs} - \pi/4)} \,, \nonumber \\
 2 \pi  f_{\rm obs} &= n \dot{\phi}_{\rm obs} (t_n), \label{eq:timefreqSPA}
\end{align}
where a dot denotes derivative with respect to time and all functions are evaluated at the stationary time $t_n$ for each $n$.
We can first use Eq.~\eqref{eq:timefreqSPA} to find
\begin{align}
 y_n (t_n) &= \left( \frac{2 \pi G M_z f_{\rm obs}(t_n)}{n c^3} \right)^{1/3} \,,
\end{align}
and then combine Eq.~\eqref{eq:SPA} with Eqs.~(\ref{eq:omegadot})--(\ref{eq:orbphase}) to obtain
\begin{align}
 \tilde{h}_{\rm obs}(f_{\rm obs}) &= \frac{G^2 M_z^2}{d_L c^5} \sum_{n \geq 1} \sqrt{\frac{5 \pi \nu}{12 n}} A_n y_n^{-7/2} e^{i \Psi_n}, \label{eq:htilde} \\
 \Psi_n &= 2 \pi f_{\rm obs} t_c - n \phi_c + \frac{3 n y_n^{-5}}{256 \nu} - \frac{\pi}{4}.
 \label{eq:Phi_constant_z}
\end{align}
For a constant redshift and at Newtonian order, this can be roughly regarded as the signal that is detected and analyzed by GW interferometers such as LISA and LIGO/Virgo.

\subsection{Time dependent redshift}

Since the expansion of the Universe and the peculiar velocity of the source are not exactly constant during the time of observation of the binary, the redshift acquires a small time-dependence. In this case, the waveform detected by an observer differs from the one derived above \cite{seto-2001,Nishizawa:2011eq,bonvin-2017}. Let us briefly repeat the derivation presented in~\cite{bonvin-2017}.
Under the assumption that both the cosmic expansion and the peculiar velocity are slowly varying over the observation time, we can linearly expand the redshift around a fixed chosen time $t_{\rm obs}^*$. As shown in Appendix~\ref{app:redshift} (see also Eq.~(36) in \cite{bonvin-2017}), the redshift at $t_{\rm obs}$ can then be written as
\begin{align}
1+z(t_{\rm obs})=(1+z_*)\Big[1+2Y(z_*)(\tau^*_{\rm obs}-\tau_{\rm obs})\Big]\, , \label{zstar}
\end{align}
where $\tau_{\rm obs} = t_c - t_{\rm obs}$ is the observed time to coalescence and $Y(z_*)$ is a small correction encoding the variation of the redshift between $t_{\rm obs}$ and $t^*_{\rm obs}$. For binaries that are observed close to the coalescence time $t_c$, a convenient choice\footnote{Both the Hubble expansion and the peculiar acceleration happen on time scales much larger than the time LISA binaries will take to coalesce. As a consequence, for the binary systems analyzed in this work, choosing $t^*_{\rm obs} = t_c$ is equivalent to choosing $t^*_{\rm obs}$ as the time when LISA starts acquiring data (see the discussion in Appendix~\ref{app:redshift}).} is to take $t^*_{\rm obs}=t_c$. In this case, Eq.~\eqref{zstar} reduces to
\begin{align}
1+z(t_{\rm obs})=(1+z_c)\Big[1-2Y_c\,\tau_{\rm obs}\Big]\, , \label{zc}
\end{align}
where
\begin{align}
&Y_c\equiv Y(t_c)=\label{eq:Ybarz}\\
&  \frac{1}{2}\left[H_{\rm obs}(t_c) - \frac{H_{\rm src}(t_c)}{1+z_c}
+ \frac{\dot{v}_{\rm src}^\shortparallel(t_c)}{c(1+z_c)} - \frac{\dot{v}_{\rm obs}^\shortparallel(t_c)}{c}\right] \,.\nonumber
\end{align}
Here $H(t)$ is the Hubble rate and $\dot{v}_{\rm src}^\shortparallel$ and $\dot{v}_{\rm obs}^\shortparallel$ are the line of sight peculiar acceleration of the source and the observer, respectively. Note that here and in the following we drop the subscript 'obs' and 'src' on the coalescence time $t_c$, to simplify the notation. Quantities at the source are always evaluated at the coalescence time in the source rest frame $t_{c\,{\rm src}}$, whereas quantities at the observer are evaluated at the coalescence time in the observer rest frame $t_{c\,{\rm obs}}$.
At zeroth order in $Y_c$ we can absorb the effect of the constant redshift $z_c$ by rewriting our expressions in terms of the redshifted mass $M_z = M(1 + z_c)$, leading again to Eqs.~\eqref{eq:htilde}--\eqref{eq:Phi_constant_z}.
In what follows we will perform all computations at linear order in $Y_c \tau_{\rm obs}\ll 1$.

We can now repeat the steps that led to Eqs.~\eqref{eq:htilde}--\eqref{eq:Phi_constant_z} for a time dependent redshift.
At the lowest order Eq.~\eqref{eq:timefreq} relates the observed time to coalescence $\tau_{\rm obs}$ to $y$.
To compute the correction at first order, we can use the frequency evolution equation in the source frame, namely Eq.~\eqref{eq:f_evol_src}, to get the evolution of the PN parameter $y$,
\begin{align}
\frac{dy}{dt_{\rm src}} &= \frac{32 \nu c^3 y^9}{5 G M} \,,
\end{align}
and then integrate to get the observed time $\tau\sub{obs}=t_c-t\sub{obs}$ as
\begin{align}
 \tau_{\rm obs} &= \int_{t_{\rm obs}}^{t_c} dt'_{\rm obs} = \int_{t_{\rm src}}^{t_c} dt'_{\rm src} (1 + z_c) (1 -2Y_c\, \tau_{\rm obs}' ) \nonumber\\
 &= \frac{5 G M_z y^{-8}}{256 \nu c^3} - \frac{25 G^2 M_z^2 Y_c y^{-16}}{65536 \nu^2 c^6} +  \ord{(Y_c\tau_{\rm obs})^2},  \label{eq:tauofy}
\end{align}
where we have inserted Eq.~\eqref{eq:timefreq} to perform the integral.
The orbital phase is again given by Eq.~\eqref{eq:orbphase}, while the observed orbital frequency becomes
\begin{align}
 \omega_{\rm obs} &= \frac{\omega_{\rm src}}{(1 + z_c) (1 -2Y_c\, \tau_{\rm obs})} \nonumber\\
 &= \frac{c^3 y^3}{G M_z} \left[ 1 +2 Y_c\,\tau_{\rm obs} \right] + \ord{(Y_c\tau_{\rm obs})^2} .
\end{align}
To compute the effect on the phase of the harmonics of the Fourier domain waveform,
we can simply write
\begin{align}
 \Psi_n &= 2 \pi f_{\rm obs} t_n - n \phi_{\rm obs} - \frac{\pi}{4} \,, \label{eq:phase_def} \\
 2 \pi f_{\rm obs} &=  n\omega_{\rm obs} (t_n) = n \frac{d\phi_{\rm obs}}{dt_{\rm obs}} (t_n) \,.
\end{align}
The time-frequency relation yields
\begin{align}
 2 \pi f_{\rm obs}
 = \frac{n c^3 y^3}{G M_z} \left( 1 + \frac{ 5 G M_z Y_c y^{-8}}{128 \nu c^3}\right) + \ord{(Y_c\tau_{\rm obs})^2} \,,
\end{align}
which inverted gives
\begin{multline}
 y^3 = \frac{2 \pi G M_z f_{\rm obs}}{n c^3} \bigg[ 1 \\
 - \frac{5 G M_z Y_c}{128 \nu c^3} \left(\frac{2 \pi G M_z f_{\rm obs}}{n c^3}\right)^{-8/3} \bigg] + \ord{(Y_c\tau_{\rm obs})^2}.
\end{multline}
Then using this relation together with Eqs.~\eqref{eq:tauofy} and~\eqref{eq:orbphase}, the phase \eqref{eq:phase_def} becomes
\begin{multline}\label{eq:pert_phase_FS} 
 \Psi_n 
 = 2 \pi f_{\rm obs} t_c - n \phi_c - \frac{\pi}{4} + \frac{3 n f_{\rm obs}^{-5/3}}{256 \nu} \left( \frac{2 \pi G M_z}{n c^3 } \right)^{-5/3} \\
 + \frac{25 n^2 Y_c f_{\rm obs}^{-13/3}}{131072 \pi \nu^2} \left( \frac{2 \pi G M_z}{n c^3} \right)^{-10/3} + \ord{(Y_c\tau_{\rm obs})^2} \,.
\end{multline}
This last equation (with $n=2$) is also derived in Appendix~\ref{app:phase}
following a slightly different route.
We can then introduce
\begin{align}
 y_f &= \left( \frac{2 \pi G M_z f_{\rm obs}}{n c^3} \right)^{1/3},
\end{align}
and we find
\begin{multline}
 \Psi_n = 2 \pi f_{\rm obs} t_c - n \phi_c - \frac{\pi}{4} + \frac{3 n}{256 \nu} y_f^{-5} \\
 + \frac{25 n G M_z Y_c}{65536 \nu^2 c^3 } y_f^{-13} + \ord{(Y_c\tau_{\rm obs})^2} \,.
 \label{eq:Psin}
\end{multline}
We can also compute the relation
\begin{align}
 y &= y_f - \frac{5 G M_z Y_c y_f^{-7}}{384 \nu c^3 } + \ord{(Y_c\tau_{\rm obs})^2}, \label{eq:yofyf}
\end{align}
from which we can infer an evolution equation for $y_f$
\begin{align}
 \frac{d y_f}{d t_{\rm obs}} &= \frac{32 \nu c^3 y_f^9}{5 G M_z} \left(1 - \frac{65 G M_z Y_c y_f^{-8}}{384 \nu c^3} \right) + \ord{(Y_c\tau_{\rm obs})^2}. \label{eq:yfdot}
\end{align}

A time dependent redshift correction appears also in the amplitude of the GW signal.
The details of this calculation are presented in Appendix~\ref{app:amplitude}.
In terms of the frequency, the waveform amplitude in the restricted waveform approximation (i.e. the leading PN order amplitude of the second harmonic $n=2$) is given by
\begin{align}
&\tilde{\mathcal{A}}_2(f_{\rm obs}) = \frac{1}{d_L(z_c)} \sqrt{\frac{5\nu}{24 \pi^{4/3} c^3}} \left( G M_{z} \right)^{5/6} f_{\rm obs}^{-7/6} \label{eq:amp_corrected_FS}\\
&\Bigg[ 1 + \frac{5\big(G M_{z}/c^3\big)^{-5/3}Y_c}{128\nu\big(\pi f_{\rm obs}\big)^{8/3}}\left(\frac{5}{2}-\frac{1}{\HH_c\chi_c} \right) \Bigg] \,,\nonumber
\end{align}
where $\chi_c\equiv c\int_0^{t_c}dt/a(t)$ denotes the background comoving distance to the source evaluated at coalescence. The luminosity distance $d_L(z)$ is affected by perturbations, namely by gravitational lensing at high redshift and by peculiar velocities at low redshift (see eq.~\eqref{eq:dL} and~\cite{Bonvin:2005ps}). 
As shown in~Appendix~\ref{app:amplitude}, these perturbations also affect the frequency-dependence of the amplitude. Indeed, the term in the square bracket of eq.~\eqref{eq:amp_corrected_FS} gets a contribution from the time variation of the luminosity distance, which is proportional to the time variation of the peculiar velocity. Note that the time variation of the gravitational lensing contribution is neglected here. This variation is proportional to the time variation of the gravitational potentials, which is expected to be much smaller than the time variation of the peculiar velocity~\footnote{In a matter dominated universe the time variation of the potentials exactly vanishes. In a $\Lambda$CDM universe this is not the case, but the variation still remains small since the potentials vary on cosmological time.}.
Eq.~\eqref{eq:amp_corrected_FS} differs from the result in~\cite{bonvin-2017}, which does not account for the time variation of the luminosity distance. For sources at low redshift, where $\chi_c$ is small, the second term is the dominant one, whereas at high redshift the first one dominates.

The expressions for the phase~\eqref{eq:Psin} and the amplitude~\eqref{eq:amp_corrected_FS} have been derived under the assumption that the scale factor $a(t)$ and the source peculiar velocity $v_{\rm src}(t)$ do not vary too much between the window of observation and the coalescence time. This is a good approximation for the binaries we will consider in the rest of this work, since the physical mechanisms producing their peculiar accelerations act on time scales much longer than the observational period of LISA, in particular on galactic time-scales. Furthermore, as we will see in Sec.~\ref{sec:results}, the binaries for which the acceleration effect is most relevant are those that are observed close to coalescence.
If other types of peculiar accelerations are analyzed, for example the perturbation of the binary's CoM due to a third close companion object, then the time variation of the acceleration must be taken into account and the analysis performed here will no longer be applicable (see e.g.~\cite{Bonetti:2017hnb,Randall:2018lnh,Robson:2018svj,Tamanini:2019awb,Wong:2019hsq,Danielski:2019rvt}).

\section{Astrophysical populations and LISA parameter estimation}
\label{sec:Fisher}

\subsection{Implementation of the peculiar acceleration effect} 

Following the definition in \cite{bonvin-2017}, we parametrize the peculiar acceleration of the binary centre of mass with a parameter $\epsilon$ such that
\begin{align}\label{eq:ep}
 \frac{\dot{v}_{\rm src}^\shortparallel(t_c)}{c(1+z_c)} = 2.4\cdot 10^{-2} \frac{H_0}{1+z_c} ~\epsilon\,.
\end{align}
This form is motivated by the particular case of a binary describing a circular orbit around the galactic center. In this case, the centripetal acceleration takes indeed the simple form
\begin{align}
 \frac{\dot{v}_{\rm src}^\shortparallel(t_c)}{c(1+z_c)} = \frac{{v}_{\rm src}^2}{r}\frac{{\bf n}\cdot{\bf e}}{c(1+z_c)} \,,
\end{align}
where $\bf e$ is the direction of the acceleration and $\bf n$ the one of the line of sight, ${v}_{\rm src}$ is the (circular) velocity and $r$ is the distance from the galaxy center.
The parameter $\epsilon$ corresponds to these quantities being normalized as follows:
 \begin{align}
 \epsilon= \left(\frac{{v}_{\rm src}}{100\,{\rm km}/{\rm sec}}\right)^2\frac{10\,{\rm kpc}}{r}  {\bf n}\cdot{\bf e}\,.
\end{align}
The argument above motivates the choice of pre-factor in Eq.~\eqref{eq:ep}; however, it is clear that $\epsilon$ can be used to parametrize any kind of peculiar acceleration, besides the centripetal one. Note that, in principle, the acceleration of the observer also plays a role (c.f.~Eq.~\eqref{eq:Ybarz}). However, we assume that this is already accounted for and subtracted via the standard GW detection procedure.

The parameter $\epsilon$ defined in \eqref{eq:ep} is convenient to parametrise the peculiar acceleration of the binary. We use it in the construction of the binary population catalogues, as presented in the following section. On the other hand, the effect on the gravitational waveform includes also the component due to the universe expansion (c.f.~section \ref{sec:theory}). 
When we implement the total effect of the centre of mass acceleration on the waveform, we do it in terms of a related parameter, $\alpha$, defined in Eq.~\eqref{eq:alphadef}. In section \ref{sub:WF_and_PE_and_bias}, we discuss the link among $\epsilon$ and $\alpha$.

\subsection{Astrophysical black hole binary populations}
\label{sub:BH_population}

We simulate two different black hole binary (BHB) populations, each assuming a different individual mass distribution: the LogFlat model assuming a uniform distribution in log-mass $dN/d m \propto m^{-1}$ for each component of the binary, and the Salpeter model assuming a Salpeter mass function $dN/dm \propto m^{-2.35}$, both with $5 M_\odot < m_i < 100 M_\odot$.
LIGO/Virgo observations provide merger rate estimates in the local Universe for both of these models~\cite{LIGOJan2017}, that we assume to be $R=32\text{ Gpc}^{-3}\text{yr}^{-1}$ for the LogFlat model, and $R=103\text{ Gpc}^{-3}\text{yr}^{-1}$ for the Salpeter model. These median rates were recently revised to be $R=19\text{ Gpc}^{-3}\text{yr}^{-1}$ for the LogFlat model, and $R=57\text{ Gpc}^{-3}\text{yr}^{-1}$ for the Salpeter model~\cite{2018arXiv181112940T}, which would roughly halve the results presented in this paper. However, the median rates used here are close to the upper limits of the confidence intervals presented in~\cite{2018arXiv181112940T}, and thus can be considered optimistic results from this point of view.

For each of those models, we simulate the local black hole binary population as follows.
We set a maximum comoving distance of $d_C\super{max}= 2$~Gpc, a minimum initial orbital frequency of $f\sub{min} = 2$~mHz, and a maximum initial orbital frequency of $f\sub{max} = 10$~Hz (the initial frequency corresponds to the frequency at which the binary is emitting when LISA starts observing). Assuming a uniform distribution of the merger times, we can compute the corresponding initial frequency evolution assuming leading order PN frequency evolution:
\begin{align}
 \frac{dN}{df} &= \frac{dt}{df} \frac{dN}{dt} = C f^{-11/3},
\end{align}
where $C$ is a constant.

We create six realisations of the binary black hole population, each for a different distribution of the acceleration parameter $\epsilon$. We use a parameter $E$ to denote the magnitude of the acceleration vector, and we create one population with $\epsilon = 0$, and one family of $\epsilon$ distributions with $E$ following a Gaussian distribution centered at $E=E_m$ with standard deviation $E_m$.
We use the following five values of $E_m = [10, 10^2, 10^3, 10^4, 10^5]$ to create 5 distributions for $E$.
These values have been chosen by equal logarithmic separation from 0 (no peculiar acceleration) to the highest expected accelerations for stellar mass BHBs living close to the galactic center ($\epsilon = 10^5$ corresponds to an acceleration of $\sim10^{-6}$ m/s$^2$).
We then obtain the corresponding $\epsilon$ as the projection of the acceleration vector along the line of sight by multiplying $E$ by a uniformly distributed number in $[-1,1]$.
In what follows we will denote these distributions as acceleration scenario 1 to 5, with scenario 0 being the distribution without any peculiar acceleration, i.e.~where $\epsilon$ has been set to zero for each event.

We then compute the yearly merger rate inside a sphere of radius $d_C\super{max}$ centered on the Solar System based on the LIGO/Virgo estimates.
For each realization, we then draw a Poisson distributed random number $N_P$ compatible with this expected rate. We then randomize the sky location and comoving distance assuming that binary systems are uniformly distributed inside the sphere, with an initial orbital frequency following a distribution proportional
to $f^{-11/3}$, and individual masses distributed by the LogFlat or the Salpeter model. We randomize all additional vector components (spins, orbital orientation) uniformly distributed on the sphere, the initial orbital phase uniformly distributed inside $[0, 2\pi]$,
and the dimensionless spin magnitudes uniformly distributed inside $[0, 1]$. 
For each binary, we compute the merger time according to the 
time-frequency relation Eq.~\eqref{eq:tauofy}.
We stop our simulation when the number of simulated systems with total mass $M < 100 M_\odot$ merging within one year, i.e. satisfying $\tau\sub{obs} < 1$~yr at the start of the mission, reaches $N_P$.

For each value of $E_m$ and for both astrophysical mass distributions, we simulate 20 catalogues taking into account two different observational scenarios, corresponding to the LISA mission taking data for 4 or 10 years.
We thus produce a grand total of 480 catalogues: 20 $\times$ 6 (values of $\epsilon$) $\times$ 2 (mass distribution) $\times$ 2 (LISA duration), spanning 3360 years worth of data.
As we will see in section~\ref{sub:detection_rates}, each catalogue contains a few hundred merging systems detectable by LISA, c.f.~Table~\ref{tab:rates}.

\subsection{Waveform generation and parameter estimation}
\label{sub:WF_and_PE_and_bias}

For the present study, we use an inspiral-only Fourier-domain gravitational waveform including spin-precession, 2.5PN order harmonics, and 3.5PN phasing~\cite{klein-2014}.
We assume zero eccentricity by considering that BBHs will have already circularized by the time they are detected by LISA, as expected for example in the classical field formation scenario \cite{Belczynski:2016obo} (see Sec.~\ref{sec:conclusions} for a discussion on eccentricity).
Note that all the quantities used in the following are in the observer frame, and thus we drop the ``obs'' subscripts from now on. We include the effects of the cosmic expansion and the peculiar acceleration by adding the linear effect in Eq.~\eqref{eq:yfdot} into the frequency evolution equation in the following way:
\begin{align}
 \dot{\omega} &= \dot{\omega}_0 \left[1 - \frac{65 \alpha}{384} \left(\frac{G M_z \omega}{c^3}\right)^{-8/3} \right], \label{eq:omegadotmod}
\end{align}
where we parametrize the acceleration effect by
\begin{align}
 \alpha &=  \frac{G M_z Y_c}{\nu c^3} \,, \label{eq:alphadef}
\end{align}
and $\dot{\omega}_0$ stands for the standard 3.5PN frequency evolution equation without acceleration effects.
Note that, neglecting the contribution of the homogeneous expansion of the Universe, which is always sub-dominant with respect to the peculiar acceleration for the systems analyzed in this work, one can relate $\alpha$ to $\epsilon$ by
\begin{equation}
	\epsilon \simeq 83.3  \frac{1+z_c}{H_0} \frac{\nu c^3}{G M_z} \alpha \qquad ({\rm no~Universe~expansion})\,.
	\label{eps_alpha}
\end{equation}
As we will see, the relative errors on $M_z$ and $\nu$ measured by LISA are always much lower than the relative error on $\alpha$, meaning that the latter dominates the relative error on $\epsilon$. We therefore use the following approximation to estimate the measurement error on $\epsilon$:
\begin{align}
\Delta \epsilon &\simeq  83.3  \frac{1+z_c}{H_0} \frac{\nu c^3}{G M_z} \Delta \alpha.
\end{align}
This approximation holds especially in the relevant cases under analysis here, i.e.~those for which the peculiar acceleration effect is sufficiently large to be measured by LISA.

Irrespective of the effect due to the modification to the phase evolution equation, the peculiar acceleration induces an additional effect on the GW amplitude, through a modification of the luminosity distance. It amounts to, as derived in Appendix~\ref{app:amplitude},
\begin{align}
\frac{d_L(z)}{d_L(z_c)} &=
1-2\left(2-\frac{1}{\HH_c\chi_c}\right)Y_c\tau_{\rm obs}\, .
\end{align}

Since every waveform amplitude $\mathcal{A}_n$ is proportional to $1/d_L$, using Eqs.~\eqref{eq:tauofy}, \eqref{eq:yofyf} and~\eqref{eq:alphadef} we multiply our waveform by a factor
\begin{align}
1 + \frac{5 \alpha}{128 y_f^8}  \left( \frac{5}{2} - \frac{1}{\HH_c \chi_c} \right).
\end{align}

The Fourier transform of the waveform is then computed using a shifted uniform asymptotic (SUA) transform~\cite{klein-2014},
\begin{align}
\tilde{h}(f) &=
\sum_n \sqrt{\frac{2 \pi}{n 
\ddot{\phi}(t_{0,n})}} \sum_{k=-k\sub{max}}^{k\sub{max}} a_{k,k\sub{max}} 
\mathcal{A}_n(t_{0,n} + k T_n) \nonumber\\
&\qquad \times e^{i[ 2 \pi f t_{0,n} - n 
\phi(t_{0,n}) - \pi/4]} \,,
\label{eq:SUAWF} \\
2 \pi f &= n 
\dot{\phi}(t_{0,n}), \\
T_n &= \frac{1}{\sqrt{n 
\ddot{\phi}(t_{0,n})}},
\end{align}
where a dot denotes differentiation with respect to time and the coefficients $a_{k,k\sub{max}}$ are constants satisfying the following linear system of equations
\begin{align}
 &\sum_{k=1}^{k\sub{max}} a_{k,k\sub{max}} + \frac{1}{2} a_{0, k\sub{max}} = 1 , \\
 &\sum_{k = 1}^{k\sub{max}} a_{k, k\sub{max}} \frac{k^{2p}}{(2p)!} = \frac{(-i)^p}{2^p p!}, \quad p \in \{1, \ldots, k\sub{max} \}, \\
 &a_{-k,k\sub{max}} = a_{k,k\sub{max}}.
\end{align}
We simulated the LISA response with a low-frequency approximation, as presented in~\cite{Cutler-1998}.

In order to evaluate the parameter estimation capabilities of LISA, we use a Fisher matrix analysis method (see e.g.~\cite{Cutler-1994,Lang-2006,Lang-2006-E1,Lang-2006-E2}). We define the inner product in the space of signals as
\begin{align}
 (a | b) &= 4 \text{Re} \int_0^\infty \frac{\tilde{a}(f) \tilde{b}^*(f)}{S_n(f)} df,
\end{align}
where $a$ and $b$ are waveform signals in the LISA detector, the tildes denote their Fourier transforms, the star denotes complex conjugation, and $S_n(f)$ is the one-sided power spectral density of the instrument's noise. We use the LISA proposal noise curve~\cite{Audley:2017drz,Babak:2017tow} with a four-year galactic binary confusion noise:
\begin{align}
 S_n(f) &= \frac{4\, S\sub{acc}(f) + S\sub{other}}{L^2} \left[ 1 + \left( \frac{2 f L}{0.41 c} \right)^2 \right] + S\sub{conf}(f),
\end{align}
with acceleration noise, noise from other sources, and confusion noise
\begin{align}
  S_{\text{acc}}(f) &= \frac{9\times 10^{-30} \text{m}^2 \text{Hz}^{-3}}{(2 \pi f)^4} \bigg[ 1 + \left(\frac{6 \times 10^{-4} \text{Hz}}{f}\right)^2 \nonumber\\
  &\times\left( 1 + \left(\frac{2.22 \times 10^{-4} \text{Hz}}{f}\right)^8 \right)  \bigg], \\
  S_{\text{other}} &= 8.899 \times 10^{-23} \text{m}^2 \text{Hz}^{-1}, \\
  S_{\text{conf}} (f) &= \frac{A}{2} e^{- s_1 f^\alpha} f^{-7/3} \left\{ 1 - \tanh \left[ s_2 (f - \kappa ) \right]\right\},
\end{align}
with $A = (3/20) 3.2665 \times 10^{-44}$~Hz$^{4/3}$, $s_1 = 3014.3$~Hz$^{-\alpha}$, $\alpha = 1.183$, $s_2 = 2957.7$~Hz$^{-1}$, and $\kappa = 2.0928 \times 10^{-3}$~Hz.

For a given binary in our catalogues, we construct its waveform as described above as a function of its parameters $h(\bm{\theta})$, where $\bm{\theta}$ is a 16-dimensional vector containing the 15 parameters describing a spinning compact object binary on a circular orbit, with the extra acceleration parameter $\alpha$. The SNR $\rho$ of this binary can be estimated as
\begin{align}
 \rho^2 &= ( h | h ).
\end{align}

The Fisher information matrix for this binary is given by
\begin{align}
 \Gamma_{ij} &= \left( \frac{\partial h}{\partial \theta^i} \right| \left.\frac{\partial h}{\partial \theta^j} \right),
\end{align}
where $\partial h/\partial \theta^i$ is the derivative of the waveform with respect to the parameter $\theta^i$.

We then construct its inverse, the covariance matrix $\Sigma = \Gamma^{-1}$. Its diagonal elements provide estimations of the statistical errors involved in the measurement of the parameters as
\begin{align}
 \Delta \theta^i &= \sqrt{\Sigma^{ii}}. \label{eq:Deltatheta}
\end{align}

We follow~\cite{Cutler-2007} to provide an estimate of the parameter estimation bias induced by neglecting the acceleration parameter $\alpha$. We define the true waveform as
\begin{align}
 h\sub{tr}(\bm{\theta}_{15}, \alpha) &= h(\bm{\theta}),
\end{align}
where $\bm{\theta}_{15}$ is the parameter vector excluding the acceleration parameter $\alpha$, and we define the approximate waveform as
\begin{align}
 h\sub{ap}(\bm{\theta}_{15}) &= h(\bm{\theta}_{15}, \alpha =0).
\end{align}

We then compute the 15-dimensional Fisher matrix $\Gamma^{(15)}$ using the approximate waveform $h\sub{ap}$:
\begin{align}
 \Gamma^{(15)}_{ij} &= \left( \frac{\partial h\sub{ap}}{\partial \theta^i_{15}} \right| \left.\frac{\partial h\sub{ap}}{\partial \theta^j_{15}} \right),
\end{align}
and its inverse $\Sigma_{15} = \left[ \Gamma^{(15)} \right]^{-1}.$

We then get an estimate of the bias that we would get in the estimation of the binary parameters by ignoring the acceleration effects as
\begin{align}
 \Delta\sub{th} \theta_{15}^i &= \Sigma_{15}^{ij} \bigg( \frac{\partial h\sub{ap} (\bm{\theta}_{15})}{\partial \theta_{15}^j} \bigg| h\sub{tr} (\bm{\theta}_{15}, \alpha) - h\sub{ap} (\bm{\theta}_{15}) \bigg).
 \label{eq:bias_def}
\end{align}
This method is valid in the high SNR limit (see~\cite{Cutler-2007} for a detailed discussion), but provides an efficient way of estimating the biases caused by a mismodelling of the intrinsic acceleration.
Note that this method is accurate only if the biases induced by the mismodelling remain sufficiently small~\cite{Cardona:2016qxn}.

\subsection{LISA mission simulations}
\label{subsubsec:simul}

For each of our 240 catalogues, we simulate two realizations of the LISA mission differing in the mission duration, that we take to be four years and ten years respectively. In each of these simulations, we compute the SNR for each binary. We separate the binaries into three distinct categories, depending on the time to merger $\tau_c$, defined as the interval of time between the beginning of LISA observation, denoted by $t_{\rm obs}^{\rm start}$, and the time of coalescence\footnote{Note that with respect to the notation in Sec.~\ref{sec:theory} here and in what follows we are using $\tau_c = \tau_{\rm obs}^{\rm start}$ to simplify the notation.}: $\tau_c\equiv t_c-t_{\rm obs}^{\rm start}$. Following~\cite{2019arXiv190511998M}, for $\tau_c < 10$~yr, we use an SNR threshold $\rho\sub{thr} = 15$ for a LISA-only detection, and we assume an archival search from a ground-based detection with SNR threshold $\rho\sub{thr} = 9.5$ for a multiband detection; for $10$~yr $< \tau_c < 100$~yr, we assume a LISA-only detection with $\rho\sub{thr} = 15$. For $\tau_c > 100$~yr, we assume a different search strategy is employed, leading to $\rho\sub{thr} = 10$: Mangiagli et al.~\cite{2019PhRvD..99f4056M} found that a Newtonian phasing will be sufficient for unbiased parameter estimation if $\tau_c > 100$~yr. We use this as evidence that those systems will be morphologically similar to Galactic white dwarf binaries, for which searches have been implemented~\cite{2010PhRvD..81f3008B,2011PhRvD..84f3009L} and found an SNR threshold of $\rho\sub{thr} \sim 10$. Note that all of these studies neglected acceleration effects. However, in a realistic scenario, non-accelerating waveforms would be used for detection and accelerating waveforms would then be used for parameter estimation. Therefore, this justifies using these studies to infer an SNR detection threshold also in the accelerating case.

For the binaries with SNR above the detection threshold, we compute the expected statistical errors $\Delta \theta^i$ on the 16 parameters, as well as the expected measurement bias $\Delta\sub{th} \theta_{15}^i$ on the 15 binary parameters excluding the acceleration effect. 
For the binaries that merge within 10 years since LISA starts taking data, we also simulate a coincident ground-based detection.
We do so by removing the entry in the Fisher matrix corresponding to the time of coalescence parameter, which we assume to be measured exactly by ground-based observations, thus reducing the dimensionality of the Fisher matrix $\Gamma$.
We therefore implicitly assume that binaries with $\rho>9.5$ in LISA are certainly detected by whatever detectors are present on Earth if they merge in a time span of ten years since the beginning of the LISA mission.

\section{Results}
\label{sec:results}

In this section we present the results of the analysis described above.
We first study for how many events the acceleration effect is detectable and then we investigate the binary parameter space to identify regions where the effect is most likely observable.
Finally we discuss whether the acceleration effect biases the measured value of the binary parameters.

\begin{table*}[!ht]
    \begin{center}
        \begin{tabular}{|c|c|ccccc|ccccc|}
            \hline
            LISA mission & Acceleration & \multicolumn{5}{c|}{LISA only}  & \multicolumn{5}{c|}{LISA+Earth ($t_c<10$y)}  \\
            \cline{3-12}
            duration & scenario & Total & 100\% & 50\% & 30\% & 10\% & Total & 100\% & 50\% & 30\% & 10\% \\
            \hline
            &
            \multirow{2}{*}{4} & 58 & 0 & 0 & 0 & 0 & 143.5 & 0 & 0 & 0 & 0 \\
            4	&			    & 40.5 & 0 & 0 & 0 & 0 & 92.5 & 0 & 0 & 0 & 0 \\
            \cline{2-12}
            years & \multirow{2}{*}{5} & 51.5 & 1 & 0 & 0 & 0 & 138 & 3 & 0 & 0 & 0 \\
            & & 38 & 0 & 0 & 0 & 0 & 92 & 1 & 0 & 0 & 0 \\
            \hline
            &
            \multirow{2}{*}{4} & 207 & 1 & 0 & 0 & 0 & 248 & 6.5 & 2 & 0 & 0 \\
            10	& 			& 157 & 0 & 0 & 0 & 0 & 159.5 & 5 & 1 & 0 & 0 \\
            \cline{2-12}
            years & \multirow{2}{*}{5} & 213. & 39.5 & 20.5 & 9.5 & 0 & 243 & 103.5 & 67 & 38.5 & 7.5 \\
            & & 154.5 & 27 & 14 & 7 & 0 & 162 & 70 & 44 & 26.5 & 5.5 \\
            \hline
        \end{tabular}
        \caption{Average (over 20 simulations) number of events for which the parameter $\alpha$ is measured with a relative error below $100\%, 50\%, 30\%$ and $10\%$.
            The upper numbers are for the LogFlat mass function, the lower ones for the Salpeter.
            ``LISA only'' denotes detections by LISA alone, while ``LISA+Earth'' are the detections for which the time of coalescence has been measured by ground-based interferometers, such as LIGO/Virgo (in this case an upper limit of 10 years has been applied on the time to coalescence from the beginning of the LISA mission). For LISA only, a detection threshold $\rho\sub{thr} = 15$ is used if $t_c < 100$~yr, and $\rho\sub{thr}=10$ otherwise; for LISA+Earth, a detection threshold of $\rho\sub{thr} = 9.5$ is used.
            }
        \label{tab:rates}
    \end{center}
\end{table*}

\subsection{Number of detections} 
\label{sub:detection_rates}

The main question that we want to address here is the following: given the astrophysical BHB populations that we produced following the method described in Sec.~\ref{sec:Fisher}, for how many events the acceleration effect will be measurable?
To answer this question we investigate the events for which the relative  uncertainty on $\alpha$, defined by $\Delta \alpha / \alpha$ where $\Delta \alpha$ is obtained through Eq.~\eqref{eq:Deltatheta}, lies below 100\%.
Note that as mentioned after Eq.~\eqref{eps_alpha} the estimated relative error on $\alpha$ can roughly be regarded as the relative error on $\epsilon$, especially when the peculiar acceleration is high so that $\Delta \alpha / \alpha<1$.

For the acceleration scenarios 0 to 3 we find no events for which $\Delta \alpha / \alpha < 1$, irrespectively of the details of the astrophysical population (Salpeter or LogFlat) or of the LISA mission duration.
This implies that the acceleration effect will not be detectable by LISA for moderate peculiar accelerations ($\epsilon \lesssim 10^3$).
This result is in agreement with previous studies \cite{bonvin-2017,Inayoshi:2017hgw} and it is here confirmed using realistic simulations of the astrophysical population of stellar-origin BHBs, over which we have imposed a distribution of moderate peculiar accelerations.
Another immediate consequence of this result is the confirmation that LISA will not be able to measure directly the expansion of the Universe.
This can be inferred because the effect is unobservable in scenario 0, where peculiar accelerations are set to zero and consequently only the cosmological acceleration is left, c.f.~Eqs.~\eqref{eq:alphadef} and \eqref{eq:Ybarz}.
This is an expected result: Ref.~\cite{Nishizawa:2011eq} demonstrated that only more advanced space-borne GW detectors such as BBO or DECIGO might be able to directly observe the cosmic expansion through its effect on the gravitational waveform.

In the remaining two scenarios, namely 4 and 5, where the acceleration vector has a typical magnitude corresponding to $\epsilon = 10^4$ and $\epsilon=10^5$ respectively, we find that the acceleration effect might be detected for some events.
From Eq.~\eqref{eq:ep} we see that these two scenarios correspond to a mean acceleration of respectively $10^{-7}$\,m/${\rm s}^2$ and $10^{-6}$\,m/${\rm s}^2$, which can typically happen if the binaries form in dense astrophysical environments, for example in nuclear star clusters or AGN disks. In the particular case of a binary system orbiting circularly around a massive central BH with mass $M_{\rm BH}$ at distance $r$, $\epsilon$ can be rewritten as
\begin{equation}
\epsilon=4.3\times 10^{-9}\frac{M_{\rm BH}}{M_\odot}\left(\frac{1 {\rm kpc}}{r} \right)^2\, .
\end{equation}
In this case, $\epsilon = 10^4$ would correspond for example to a binary system orbiting at $66$\,pc of a BH of mass $M_{\rm BH}=10^{10}M_\odot$ (whose Schwarzschild radius is $r_s=9.4\times 10^{-4}$\,pc).

In Table~\ref{tab:rates} we report the total number of LISA detections together with the number of events for which the estimated error on $\alpha$ is measured with a relative uncertainty below 100\%, 50\%, 30\% and 10\%.
Results are shown for both a 4 years and 10 years LISA mission and for both BH populations: LogFlat (upper numbers in each line) and Salpeter (lower numbers in each line).
We report detections by LISA only and for multiband events detected first by LISA and then by ground-based detectors such as LIGO and Virgo. For the latter class of detections, we select events with a time 
 to coalescence (measured from the start of the LISA mission) below 10 years.
The detection by ground-based interferometers is accounted for by fixing the value of the coalescence time $t_c$ to its true value, and it is practically implemented by eliminating the row and column corresponding to the parameter $t_c$ in the Fisher matrix.

From Table~\ref{tab:rates} it is clear that even for scenarios 4 and 5 the acceleration effect will not be measurable if we assume that the LISA mission will last 4 years.
Only very rare events will have a peculiar acceleration strong enough to be barely detectable, implying that no definitive conclusions on the astrophysical processes of BHB formations will be extracted.
The same conclusions can be drawn for the scenario 4 with a 10 years LISA mission, where the effect can be measured for few events only if a coincident detection with ground-based interferometers is made.
Scenario 4 yields thus results similar to scenarios 0 to 3 where the number of events for which the acceleration effect is measurable is too low to extract any useful information on the BHB population.

The situation is different for scenario 5 if a 10 year LISA mission is considered.
According to Table~\ref{tab:rates} LISA will be able to measure (with $\Delta \alpha / \alpha < 1$) the acceleration effect in about $17$\% of the total detected events, and for more than about $4$\% of them (roughly $\sim 10$ events) the peculiar acceleration will be determined with a relative precision better than 30\%.
These results drastically improve for multiband coincident detections: more than 40\% of these events will present measurable peculiar accelerations, with about 16\% and 3\% of them presenting precisions better than 30\% and 10\%, respectively.
Moreover, these percentages are independent of the BH population considered, since they roughly hold for both the Salpeter and LogFlat mass functions.
Table~\ref{tab:rates} shows, however, that different choices for the BHB population model do influence the total number of events for which the acceleration is detectable, since the LogFlat mass function provides systematically higher event numbers, due to LISA being more sensitive to higher mass systems~\cite{Sesana:2016ljz}.
Note that the total number of detections quoted here are higher than estimated in some other works (e.g.~\cite{2019arXiv190511998M}). This is due to our choice of a higher mass limit of $100 M_\odot$ instead of $50 M_\odot$.

These results show that a BHB population with high enough peculiar accelerations will imprint a distinguishable signature on the events detected by LISA only if the mission will reach its maximum possible duration.
Such a population would be well representative of a characteristic BH formation channel where binaries are created in AGN disks close to the galactic center \cite{Inayoshi:2017hgw}.
The results reported here thus support the science case for a LISA mission duration longer than the nominal 4 years period currently planned \cite{Audley:2017drz}.
If LISA observes GWs somewhat continuously for almost 10 years, then we will have the chance to gather useful information on possible formation channels of BHBs as well as their distance from the center of the hosting galaxy (for BHBs orbiting a galaxy the peculiar acceleration is expected to be roughly proportional to the inverse of the distance from the galactic center, assuming simple circular motion).

\subsection{Effect across the parameter space} 
\label{sub:parameter_estimation}

\begin{figure*}
	\includegraphics[width=\textwidth]{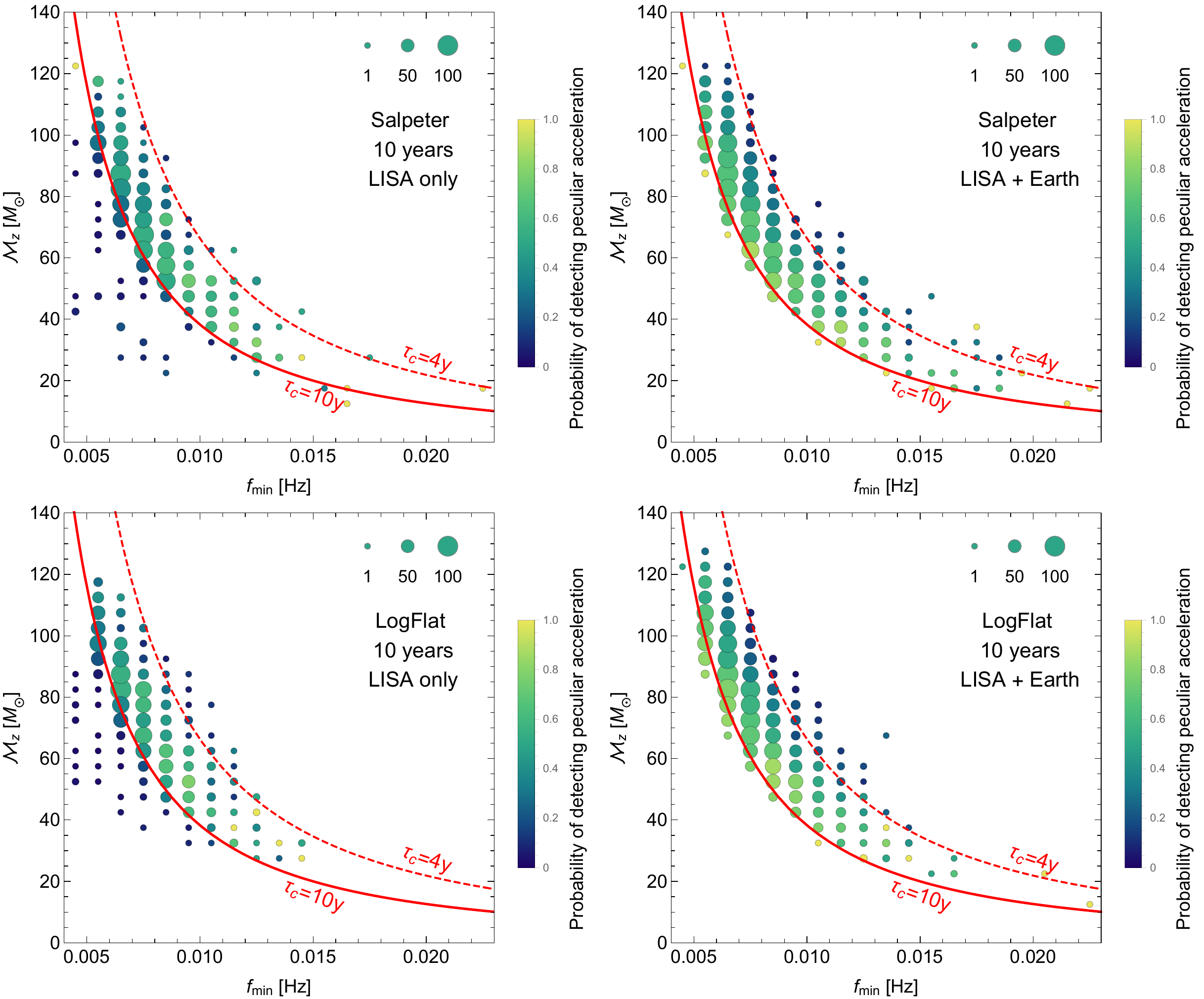}
	\caption{
	Probability of detecting the acceleration effect (i.e.~the ratio among the number of events with $\Delta\alpha/\alpha<1$ and the total number of detected events) in the $(f_{\rm min}, \mathcal{M}_z)$-parameter space, where $f_{\rm min}$ is the initial frequency and $\mathcal{M}_z$ the redshifted chirp mass of the BHBs.
	The size of each bubble corresponds to the total number of events in the respective 2D bin of 1 mHz times 5 $M_\odot$ for which the peculiar acceleration effects is detected.
	We have chosen 10 years of LISA mission duration, and selected the acceleration scenario 5.
	Binaries at the left of the solid-red curve merge in more than 10 years (from the start of LISA observations) and are therefore excluded in the joint LISA + Earth-based detections.
	} 
	\label{fig:prob_fmin_Mc}
\end{figure*}

The magnitude of the acceleration effect on the gravitational waveform depends on the BHB parameters: there are regions in the binary parameter space where the acceleration effect is stronger, and thus more likely to be detected.
For example, from Eq.~\eqref{eq:omegadotmod} one can roughly infer that, for the same value of $\alpha$, the product $M_z\omega$ should be maximized (c.f.~Fig.~\ref{fig:prob_fmin_Mc}).
The question we want to address in this section is the following: for which values of the BHB parameters can we expect to measure the acceleration effect?
In order to answer this question we focus on the only scenario where a sufficiently large number of detections is expected, which, according to Sec.~\ref{sub:detection_rates}, corresponds to scenario 5 with a LISA mission of 10 years.
However, irrespective of the LISA mission duration or the BH formation scenario, this analysis allows us to select, in a set of measured events, those for which it would be more plausible to detect the center of mass acceleration.

We focus our analysis on two parameters in particular: $\mathcal{M}_z = M_z \nu^{3/5}$ (the redshifted chirp mass), and $f_{\rm min}$ (the GW frequency at the initial time of LISA observations).
All other physically interesting parameters follow distributions roughly correlated with the SNR: the acceleration effect is better measured at lower distances, for symmetric mass ratios close to $\nu = 1/4$ and for higher values of $\alpha$.
In Fig.~\ref{fig:prob_fmin_Mc} we show the probability of measuring the acceleration effect in the $(f_{\rm min}, \mathcal{M}_z)$-parameter space, where for each 2D bin of 1mHz times 5 $M_\odot$ this probability is defined as the ratio between the number of events for which the relative $1\sigma$ error on $\alpha$ is below 100\% and the total number of detections in that bin.
In order to provide additional information on how confident the statistics of each 2D bin is, bins are represented as spherical bubbles whose sizes are proportional to the number of events for which the peculiar acceleration is detected.
Clearly, bins with a higher number of events are statistically more reliable.
We report results for both mass functions (Salpeter and LogFlat) and with both LISA only detections and multiband detections involving ground-based observations.
In each panel we also plot the lines to the right of which binaries will merge in less than 10 or 4 years after LISA starts taking data.

It is evident from Fig.~\ref{fig:prob_fmin_Mc} that the acceleration effect is measurable preferably for  binaries with $\tau_c \lesssim 10$ years.
Note that multiband events are limited by the condition $\tau_c < 10$ years, so no events appear below the solid red line in the right panels of Fig.~\ref{fig:prob_fmin_Mc}. We therefore find that optimal detection is obtained for times to coalescence close to the mission duration. This confirms what was obtained by previous studies (cf.~Fig.~3 of \cite{Inayoshi:2017hgw}), which, however, did not consider simulations of realistic astrophysical BHB populations.
This result can be understood by two balancing effects: on the one hand, the parameters of the binaries can be better measured when we are observing close to coalescence (very far from coalescence the waveform evolves indeed very slowly inducing strong degeneracies between the parameters of our waveform model).
On the other hand, the dephasing generated by the binaries' acceleration accumulates over the time of observation, and therefore the longer we observe, the better $\alpha$ can be determined.
The optimal situation to detect the acceleration effect is to observe almost up to coalescence.

We also plot on Fig.~\ref{fig:prob_fmin_Mc} the curve $\tau_c=4$ years (dashed-red), corresponding to binaries which coalesce 4 years after the beginning of observation. 
This curve roughly marks the upper edge of the parameter space region for which the acceleration becomes measurable for a consistent number of events with a LISA mission lasting 10 years.
This explains why, with a 4 years mission, the acceleration effect is not detectable for many events: we need to observe a binary for more than 4 years before coalescence in order to reach a non-negligible detection probability.
This plot also suggests that increasing the duration of the mission slightly above 4 years would already allow us to detect the acceleration effect with a 20-30 percent probability.

The results we obtain here, shown in Fig.~\ref{fig:prob_fmin_Mc}, are coherent with what found in \cite{Inayoshi:2017hgw}, despite the differences between the two analyses: in \cite{Inayoshi:2017hgw} there was no simulated binary population and the mass ratio and the redshift were fixed, furthermore both the waveforms and the LISA  mission characteristics were different.   
The upper limit around redshifted chirp masses of 120 $M_\odot$
and the lower limit around $\mathcal{M}_z \simeq 10-20 M_\odot$ are both defined by the lack of BHBs with higher and lower masses, respectively, in our population of events detected by LISA (the maximum chirp mass for our population is $87 M_\odot$ in the source frame and $\sim 130 M_\odot$ in the detector frame). 
It appears from our analysis that the peculiar acceleration is comparatively easier to measure for low mass binaries, in particular for LISA alone. 
We also observe that accounting for multiband observations substantially improves the detection probability.
Note however that, due to the low number of events detected by LISA in a single mission lifetime in the region $\mathcal{M}_z \simeq 20-40 M_\odot$, $f_{\rm min} \gtrsim 12$ mHz, more statistics (i.e.~more simulated populations of BHBs) would be required in order to better characterize this region.
Nevertheless, our results indicate that, when the real data is available, it will be worthwhile to check for the presence of peculiar acceleration in events having binary parameters close to the above mentioned region, as they are more likely to be affected by the acceleration effect.

\subsection{Biases} 
\label{sub:biases}

We finally investigate the bias induced by the acceleration effect in case it is neglected in the parameter estimation analysis.
The questions we want to address here are the following: how many events will be biased by the acceleration effect if its contribution to the GW waveform is ignored? Which waveform parameters will be biased the most and for which values?
To address these questions we estimate the bias on each waveform parameter of each event using the bias estimator discussed in Sec.~\ref{sec:Fisher}, namely Eq.~\eqref{eq:bias_def}.
We analyze the results derived with LISA in this section and we do not consider coincident detections with Earth-based interferometers.
We are in fact interested in understanding how LISA parameter estimation might be biased by the peculiar acceleration effect and how this might affect the expected coincident detection with ground-based instruments (e.g.~with regards to the expected time of coalescence or sky position).

\begin{table}[t!]
\begin{center}
\begin{tabular}{|c|C{0.5cm}C{0.5cm}C{0.55cm}C{0.55cm}C{0.65cm}|C{0.5cm}C{0.55cm}C{0.55cm}C{0.55cm}C{0.65cm}|}
\hline
  & \multicolumn{5}{c|}{4 years LISA} & \multicolumn{5}{c|}{10 years LISA} \\
\hline
 Waveform & \multicolumn{5}{c|}{Acceleration scenario} & \multicolumn{5}{c|}{Acceleration scenario}  \\
 parameter & 1 & 2 & 3 & 4 & 5 & 1 & 2 & 3 & 4 & 5 \\
\hline
 \multirow{2}{*}{$M_1$} & 0. & 0.1 & 4.1 & 11.6 & 25.7 & 0.6 & 2.9 & 12.3 & 30.4 & 41.2 \\
    				    & 0. & 0. & 4.1 & 9.6 & 21 & 0.5 & 2.3 & 12.8 & 30.3 & 38.7 \\
\hline
 \multirow{2}{*}{$M_2$} & 0. & 0.1 & 2.9 & 10.1 & 24.6 & 0.2 & 2. & 10.5 & 30.3 & 41.3 \\
    				    & 0. & 0. & 2.7 & 8.7 & 19.4 & 0.2 & 1.4 & 11. & 30. & 39.1 \\
\hline
 \multirow{2}{*}{$\cos(\theta_N)$} & 0. & 0. & 0. & 1.1 & 10 & 0. & 0. & 1. & 11.6 & 23.7 \\
 								   & 0. & 0. & 0. & 1.2 & 10.7 & 0. & 0. & 1.1 & 11.7 & 23.4 \\
\hline
 \multirow{2}{*}{$\phi_N$} & 0. & 0. & 0. & 0.3 & 10 & 0. & 0. & 0.4 & 11.7 & 29.8 \\
 						   & 0. & 0. & 0. & 0.4 & 10.4 & 0. & 0. & 0.2 & 10.4 & 30. \\
\hline
 \multirow{2}{*}{$d_L$} & 0. & 0. & 0.2 & 5.5 & \textbf{54.7} & 0. & 0. & 4.7 & 46.5 & \textbf{91.4} \\
    				    & 0. & 0. & 0.1 & 4.4 & \textbf{48.4} & 0. & 0. & 4.5 & 43.5 & \textbf{90.3} \\
\hline
 \multirow{2}{*}{$cos(\theta_L)$} & 0. & 0. & 0.3 & 6.4 & 34.8 & 0. & 0.3 & 6.2 & 34.4 & 63.2 \\
 								  & 0. & 0. & 0.5 & 4.4 & 32.9 & 0. & 0.3 & 5.7 & 32. & 61.5 \\
\hline
 \multirow{2}{*}{$\phi_L$} & 0. & 0. & 1. & 7.7 & 38.4 & 0.1 & 0.8 & 8.4 & 36. & 65 \\
 						   & 0. & 0. & 1.1 & 6.3 & 33.9 & 0.1 & 0.9 & 7.4 & 35.4 & 64.6 \\
\hline
 \multirow{2}{*}{$t_c$} & 0. & 1.4 & 11.1 & 20.4 & 32.1 & 0.1 & 3.1 & 16.5 & 31. & 37 \\
 						& 0. & 0.9 & 12.2 & 21.1 & 29.2 & 0.2 & 2.5 & 16.1 & 31.1 & 37 \\
\hline
 \multirow{2}{*}{$\phi_c$} & 0. & 1.2 & 9.2 & 20.2 & 33.1 & 0.1 & 2.4 & 14.7 & 30.4 & 39.9 \\
 						   & 0. & 0.9 & 10.6 & 20.9 & 30.4 & 0.1 & 2.1 & 14.9 & 31.3 & 38.2 \\
\hline
 \multirow{2}{*}{$cos(\theta_1)$} & 0.1 & 0.6 & 1. & 5.9 & 17.5 & 1.2 & 3.3 & 9.8 & 23.7 & 35.9 \\
 								  & 0. & 0.3 & 1.2 & 3.6 & 14.3 & 1.1 & 3.4 & 8.8 & 22.5 & 35.7 \\
\hline
 \multirow{2}{*}{$\phi_1$} & 0. & 0. & 0.4 & 3.2 & 17.3 & 0.1 & 0.5 & 5.2 & 21.5 & 36.3 \\
 						   & 0. & 0. & 0.1 & 2.3 & 12.8 & 0. & 0.6 & 3.8 & 19.9 & 34.1 \\
\hline
 \multirow{2}{*}{$\chi_1$} & 0. & 0. & 0.1 & 0.9 & 15.6 & 0. & 0.1 & 1.8 & 17.6 & 36.8 \\
 						   & 0. & 0. & 0. & 0.7 & 13.9 & 0. & 0.1 & 1.6 & 17.5 & 37.3 \\
\hline
 \multirow{2}{*}{$cos(\theta_2)$} & 0. & 1. & 2.5 & 6.8 & 18.8 & 1.6 & 5.9 & 13.3 & 26. & 36.2 \\
 								  & 0. & 0.6 & 2.2 & 4.4 & 17.3 & 1.5 & 5.1 & 11.7 & 23.7 & 36 \\
\hline
 \multirow{2}{*}{$\phi_2$} & 0. & 0.1 & 1.1 & 4. & 17.2 & 0.4 & 1.5 & 7.1 & 21.6 & 33.9 \\
 						   & 0. & 0.2 & 1. & 3.1 & 15.3 & 0.3 & 1.7 & 6.3 & 21.3 & 33.5 \\
\hline
 \multirow{2}{*}{$\chi_2$} & 0. & 0. & 0. & 1.3 & 12.8 & 0. & 0.1 & 2. & 18. & 34.4 \\
 						   & 0. & 0. & 0.1 & 1.5 & 13.1 & 0. & 0.1 & 1.9 & 17.2 & 34.7 \\
\hline
 \multirow{2}{*}{\bf any} & 0.1 & 3.4 & 16.8 & 38.2 & 76.9 & 3.9 & 14.5 & 42.6 & 77.8 & 96.9 \\
					   		& 0. & 2. & 17.8 & 35.7 & 76.2 & 3.6 & 13.7 & 41. & 76.7 & 96.2 \\
\hline
\end{tabular}
\caption{Percentage probability of being biased for each waveform parameter. This probability is defined as the fraction of events for which the bias on a given parameter is larger than the statistical 1$\sigma$ error on that parameter, with respect to the total number of events. Upper numbers correspond to the LogFlat mass function and lower numbers to the Salpeter one. $M_1$ and $M_2$ represent the individual masses, $\theta_N$ and $\phi_N$ the spherical angles of the line-of-sight vector, $d_L$ the luminosity distance, $\theta_L$ and $\phi_L$ the spherical angles of the orbital angular momentum, $t_c$ the coalescence time, $\phi_c$ the coalescence orbital phase, $\theta_1$ and $\phi_1$ the spherical angles of the initial spin of the body of mass $M_1$, $\chi_1$ its dimensionless magnitude, and $\theta_2$, $\phi_2$, and $\chi_2$ the corresponding parameters for the spin of the body of mass $M_2$. All vector angles are defined in an inertial frame tied to the ecliptic. The last line corresponds to the percentage of systems for which at least one of the parameters is biased.}
\label{tab:biases}
\end{center}
\end{table}

In Table~\ref{tab:biases} we report the percentage of biased events for the scenarios with non-zero peculiar acceleration. We average over the 20 realizations we have constructed, considering a LISA mission durations of 4 and 10 years, the acceleration scenarios 1 to 5, and both BH mass functions: LogFlat (upper numbers) and Salpeter (lower numbers in each line).
For each waveform parameter an event is considered biased if the corresponding bias is greater than the statistical 1$\sigma$ error derived from the Fisher analysis without the acceleration effect, i.e.~if the bias $\Delta\sub{th} \theta_{15}^i$ is larger than the error bar $\Delta\theta^i_{15}$ (see Sec.~\ref{sub:WF_and_PE_and_bias}).
The percentage of biased events at the bottom of Table~\ref{tab:biases} is defined as the number of events for which at least one parameter is biased over the total number of events.

From Table~\ref{tab:biases} we first notice that for each parameter the fraction of biased events is roughly the same for both BH populations (the Salpeter mass function systematically provides slightly lower percentages).
This implies that the bias results of our analysis are roughly independent of the details of the underlying BHB astrophysical population.
We recall though that the number of detections for the LogFlat mass function is higher than the Salpeter case (cf.~Table~\ref{tab:rates}), meaning that in the latter case we would expect a lower absolute number of biased events.

From the last row of Table~\ref{tab:biases} it is furthermore clear that few if any events will be biased in acceleration scenario 1, irrespectively of the BH mass function (Salpeter or LogFlat) and of the LISA mission duration.
The situation changes for acceleration scenarios with higher peculiar accelerations.
For the acceleration scenario 2, we find that $\sim 14\%$ of events will be biased if the LISA mission lasts 10 years, and a negligible amount if it lasts 4 years. 
For acceleration scenario 3 we find that $\sim 40\%$ of the events are biased if the LISA mission lasts 10 years, and about $16\%$ if it lasts 4 years. 
For acceleration scenarios 4 and 5, most events will be biased if the mission lasts 10 years, up to $\sim 96\%$ for scenario 5.  

From Table~\ref{tab:rates} we learned that a non-negligible fraction of events of scenario 5 with a 10-year LISA presents measurable peculiar acceleration. Therefore, finding many biased events in this case is expected.
The acceleration effect will be measurable for some of the biased events, meaning that, for these, the bias on the waveform parameters may be corrected using the right, accelerated waveform.
Somehow more surprising is the large number of biased events obtained in scenarios with smaller acceleration.
According to Table~\ref{tab:rates}, only very few events have measurable peculiar acceleration in these cases, while here we find that a significant fraction of events (reaching $\sim75\%$ for acceleration scenario 5 and a 4-year mission, and scenario 4 and a 10-year mission), at least one of the waveform parameters is biased.
One possible explanation is that the parameter $\alpha$ is degenerate with the other parameters.
In this case, $\alpha$ is not large enough to break this degeneracy and allow for a detection of the effect.
However, when $\alpha$ is not included in the modelling of the signal, the effect is reabsorbed into a shift in the other parameters, particularly the ones that are degenerated with $\alpha$. 
As a consequence, in this scenario the acceleration effect is not detectable, but the bias induced on the other parameters can be important.

It is particularly important to remark that the parameter estimation of BHBs formed in AGN disks (acceleration scenario 5) might be biased by the peculiar acceleration effect even for the LISA nominal mission duration of 4 years.
Note that in this case, adding $\alpha$ to the modelling of the signal would remove the bias on the other parameters.
This is the case even for the events for which $\alpha$ is not measurable, at the expense of increasing the other parameters errors.

\begin{table}[t!]
\begin{center}
\begin{tabular}{|c|C{0.5cm}C{0.5cm}C{0.5cm}C{0.7cm}C{0.7cm}|C{0.5cm}C{0.5cm}C{0.7cm}C{0.7cm}C{0.7cm}|}
\hline
  & \multicolumn{5}{c|}{4 years LISA} & \multicolumn{5}{c|}{10 years LISA} \\
\hline
 Bias & \multicolumn{5}{c|}{Acceleration scenario} & \multicolumn{5}{c|}{Acceleration scenario}  \\
 on $d_L$ & 1 & 2 & 3 & 4 & 5 & 1 & 2 & 3 & 4 & 5 \\
\hline
 \multirow{2}{*}{$>1\sigma$} & 0. & 0. & 0.2 & 5.5 & 54.7 & 0 & 0 & 4.6 & 46.1 & 90.8 \\
    				    	 & 0. & 0. & 0.1 & 4.4 & 48.4 & 0 & 0 & 4.2 & 44.9 & 90 \\
\hline
 \multirow{2}{*}{$>2\sigma$} & 0. & 0. & 0. & 2.7 & 41.6 & 0 & 0 & 2.1 & 34.9 & 86.3 \\
    				    	 & 0. & 0. & 0. & 2.3 & 36.1 & 0 & 0 & 2.2 & 33.3 & 85.4 \\
\hline
 \multirow{2}{*}{$>3\sigma$} & 0. & 0. & 0. & 1.5 & 31.6 & 0 & 0 & 1.2 & 27.4 & 81.6 \\
 							 & 0. & 0. & 0. & 1.5 & 27.4 & 0 & 0 & 1.3 & 26.2 & 80.8 \\
\hline
\end{tabular}
\caption{Percentage of events for which the bias on the luminosity distance $d_L$ is bigger than its corresponding 1$\sigma$, 2$\sigma$ and 3$\sigma$ uncertainties.
Upper numbers are for LogFlat and lower numbers are for Salpeter mass function.
}
\label{tab:bias_dL}
\end{center}
\end{table}

We can now look at which waveform parameters are more biased on average.
From Table~\ref{tab:biases} it appears that the parameter which is biased in the highest percentage of cases is the luminosity distance $d_L$.
We point out that this result is particularly important for LISA applications of stellar mass BHBs as standard sirens \cite{DelPozzo:2017kme,Kyutoku:2016zxn}, since a possible systematic error on $d_L$ will negatively affect the accuracy with which cosmological parameters can be measured.
In fact since at low redshift 
\begin{equation}
d_L(z_c)\simeq \frac{c}{H_0} z_c \,,
\end{equation}
a bias of 1$\sigma$ on $d_L$ directly translates into a bias of 1$\sigma$ on the Hubble parameter $H_0$.
In Table~\ref{tab:bias_dL}, we show the percentage of events for which the bias on the luminosity distance is larger than the 1, 2 and 3$\sigma$ error on $d_L$, respectively.
We see that more than $\sim$30\% of events will present a bias bigger than 3$\sigma$ in the acceleration scenario 5 for 4 years of LISA observations. Similarly, in the acceleration scenario 4 for 10 years of LISA observations we find that this number is $\sim$25\%.
This implies that in these scenarios any measurement of $H_0$ can be strongly affected by the acceleration effect, introducing a systematic error in low-redshift cosmological measurement with LISA.
The situation is even worse in the acceleration scenario 5 for 10 years of LISA measurement, where up to $\sim$80\% of events will present a bias in $d_L$ higher than its 3$\sigma$ error.
The solution to avoid this bias is then to introduce the acceleration effect in the modelling of the waveform.
Even if the acceleration parameter $\alpha$ is not large enough to be detected, by adding its effect to the waveform we ensure that the other parameters, including the luminosity distance, are not biased.
This of course comes at the cost of widening the error bars, especially if $\alpha$ is degenerated with other parameters.
Fortunately the sky localization of the source, represented by the angles $\theta_N$ and $\phi_N$, seems not to be substantially biased, except in the most extreme case (10-year mission with acceleration scenario 5), where we expect to detect the acceleration effect for a large number of events (cf.~Table~\ref{tab:rates}).
This is good news not only for standard sirens\footnote{A good sky localization is required to apply the so-called ``statistical'' method for standard sirens, where the sky position of GW sources is cross-correlated with galaxy catalogues \cite{Schutz:1986gp,DelPozzo:2011yh,Fishbach:2018gjp}.}, but also for multiband and multi-messenger detections (although counterparts are not really expected for BHB).

Similarly to what happens for the luminosity distance, the direction of the orbital angular momentum of the binary, represented by the angles $\theta_L$ and $\phi_L$, is also biased for a large fraction of events in the highest acceleration scenarios. On the other hand, the bias on the time of coalescence $t_c$ and the phase at coalescence $\phi_c$ appears at substantially lower values of the peculiar acceleration. Small fractions of events are biased even for acceleration scenarios 2 and 3. These two parameters seem therefore to be particularly sensitive to the acceleration effect, a result to which~\cite{bonvin-2017} already hinted\footnote{Note, however, that the analysis of~\cite{bonvin-2017} concentrated on very low-mass binaries, and on the probability of recovering $t_c$ in less than one minute. Since low-mass binaries are extremely far from coalescence, big absolute errors on $t_c$ are not at all surprising.}. A possible interpretation of this fact is that $t_c$ and $\phi_c$ are more degenerate with $\alpha$ than the other binary parameters.

\begin{figure*}
	\includegraphics[width=\textwidth]{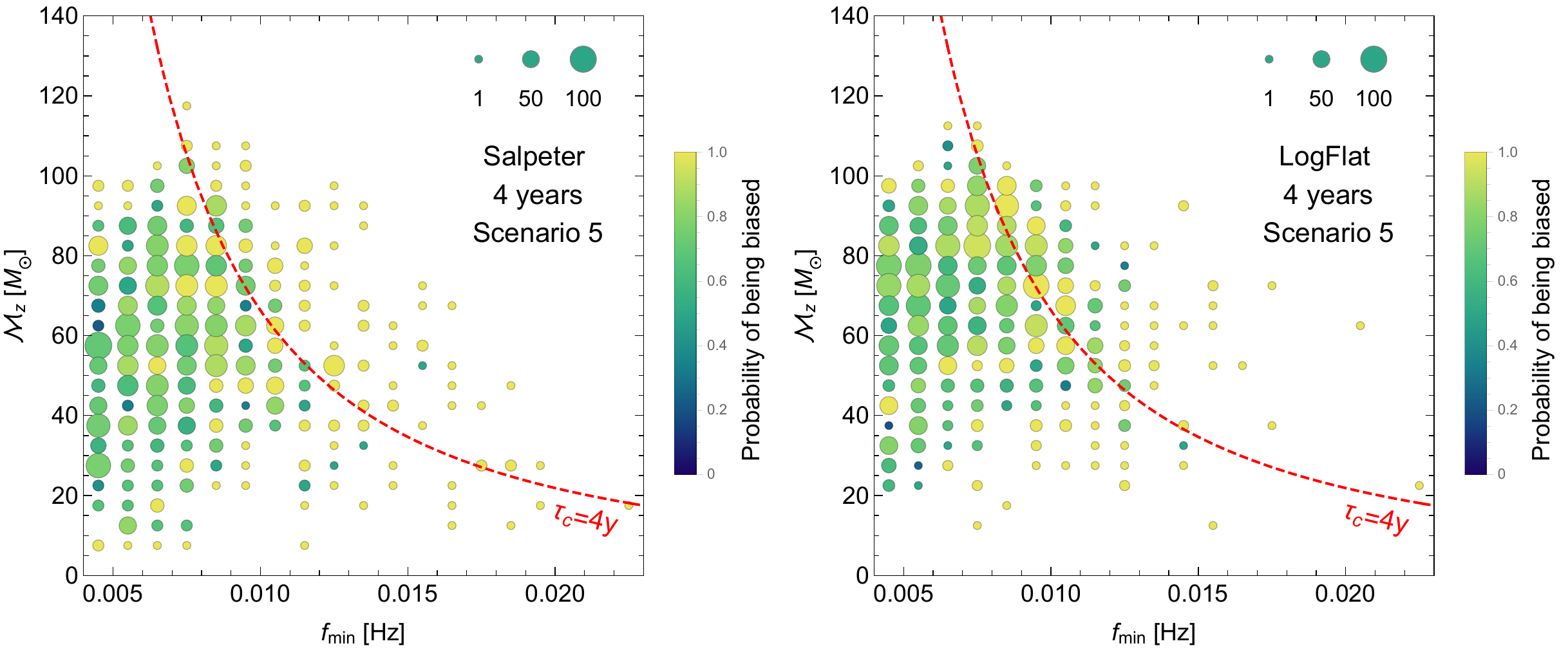}
	\caption{Probability of being biased by the acceleration effect in the $(f_{\rm min}, \mathcal{M}_z)$-parameter space, where $f_{\rm min}$ is the initial frequency and $\mathcal{M}_z$ the redshifted chirp mass of the BHBs.
	The size of each bubble corresponds to the total number of biased events in the respective 2D bin of 1 mHz times 5 $M_\odot$.
	We have chosen 4 years of LISA mission duration, and selected the acceleration scenario 5. Binaries at the left of the dashed-red curve merge in more than 4 years.
	} 
	\label{fig:bias_fmin_Mc}
\end{figure*}

We can now turn our attention to the parameter space, and investigate for which values of the binary's parameter a high fraction of detected events is biased.
We again focus only on the parameters $\mathcal{M}_{z}$ (redshifted chirp mass) and $f_{\rm min}$ (initial frequency), to better compare with the analysis of Sec.~\ref{sub:parameter_estimation}.
In Fig.~\ref{fig:bias_fmin_Mc}, we plot the probability of being biased, defined in each 2D bin as the ratio between the total number of biased events (those with bias larger than the 1$\sigma$ error, as done for Table~\ref{tab:biases}) and the total events in the bin, in the ($\mathcal{M}_z$--$f_{\rm min}$) parameter space.
Again, in order to provide additional information on how confident the statistics of each 2D bin is, bins are represented as spherical bubbles whose sizes are proportional to the number of biased events.
Thus, bins with a higher number of events are statistically more reliable.
We only show the result for acceleration scenario 5, a 4-year LISA, and both Salpeter and LogFlat mass functions: this represents the most interesting case for a LISA mission of nominal duration.
Indeed, the same figure in the case of a 10-year LISA mnission is less informative, as the results are biased across almost all the parameter space (cf.~the numbers in Table~\ref{tab:biases}).

Fig.~\ref{fig:bias_fmin_Mc} shows that events with the time to coalescence close to and lower than the mission duration have higher probability of being biased. In agreement with the results of Sec.~\ref{sub:parameter_estimation}, for these events we expect relevant systematic effects induced by the peculiar acceleration. Moreover, as already hinted by Table~\ref{tab:biases}, many events for which the acceleration effect is not measurable turn out to be biased: e.g.~events with $\tau_c$ larger than the mission lifetime, as shown in Fig.~\ref{fig:bias_fmin_Mc}. Of particular interest for multiband detections are low mass events, as well as events with $\tau_c$ shorter than the mission duration (events lying above the red curve in Fig.~\ref{fig:bias_fmin_Mc}), as they merge at higher frequencies and in a relatively short time.
These events appear to have a higher probability of being biased.


\subsection{Intermediate mass black holes} 
\label{sub:intermediate_mass_black_holes}

We have also checked the detectability of the peculiar acceleration for intermediate mass BHBs (IMBHBs).
By generalizing the procedure described above, we have constructed a population of IMBHBs with single BH masses up to 500 $M_\odot$.
Since the rates of IMBHBs are presently unknown and poorly constrained, we have not used the normalization of such a population for our analysis, and we have abstained from making any prediction regarding the number of detections that LISA could attain.
The main finding of this analysis is that, in analogy with stellar-origin BHBs, IMBHBs whose time to coalescence at the beginning of observations matches the mission duration, present a higher probability of yielding a measurable peculiar acceleration effect.
Although the merger rate of an intermediate mass population of BHBs is completely unknown, it is interesting to note that they might actually be efficiently created in AGN disks \cite{McKernan:2012rf,McKernan:2014oxa,Bellovary:2015ifg}, namely our scenario 5.
The astrophysical models leading to IMBHs in AGN disks may thus be efficiently tested by LISA, which will certainly detect these systems and measure their peculiar acceleration if the mission lifetime will be extended over its nominal 4 year duration.

\section{Discussion and conclusion}
\label{sec:conclusions}

The present work has been dedicated to assess the potential of LISA to detect the peculiar acceleration of stellar-origin BHBs.
Based on the idea that the peculiar acceleration of a binary alters its gravitational waveform~\cite{bonvin-2017}, we investigated how well it can be measured, using simulated BHB populations.

Our results confirm that the BHBs which are more likely to present a measurable peculiar acceleration are those that will merge close to the end of the mission~\cite{Inayoshi:2017hgw}. The reason for this is twofold: on the one hand, the dephasing due to the peculiar acceleration accumulates with the time of observation; on the other hand, the uncertainty on the other waveform parameters decreases when observed close to coalescence, as many degeneracies in the phase are broken by the fast frequency evolution of the waveform.

The systems for which a peculiar acceleration can be measured are usually those with high SNR (recall that SNR $\propto f^{2/3}$).
The Fisher matrix approach employed here is expected to provide a sufficiently good approximation for the true posteriors of high SNR binaries.
Therefore, the BBHs for which the peculiar acceleration is measurable should be well described by our analysis.
Moreover, a full Bayesian estimation for similar systems detectable by LISA, namely inspiraling binaries moving around a third body, has already been performed in the literature~\cite{Robson:2018svj}, showing that the Fisher matrix approach provides a reasonable approximation (see also \cite{Caputo:2020irr}).

We find that with the nominal LISA mission lifetime (4 years), no significant measurement is possible, irrespective of the magnitude of the peculiar acceleration considered (within plausible scenarios).
On the other hand, we show that a LISA mission lasting 10 years will be able to detect the peculiar acceleration of a relevant fraction of binaries, if these form in dense astrophysical environments, for example in nuclear star clusters or AGN disks, where peculiar accelerations might reach values of ${\sim}10^{-6}$ m/s$^2$ (corresponding to $\epsilon\sim 10^5$; cf.~Eq.~\eqref{eq:ep}). 
Our results also suggest that, increasing the mission duration to little over 4 years might already be enough to discriminate between different BHB formation channels, shedding new light on the processes at work when BHBs are created.

The peculiar acceleration effect investigated here will moreover be complementary to other effects expected from stellar BHBs in close orbit around a massive BH, in particular large eccentricities up to merger due to Eccentric Kozai-Lidov evolution \cite{Hoang:2019kye}.
In fact, LISA could detect eccentric stellar-mass BBHs if they interact with other close-by massive bodies \cite{Seto:2016wom,Nishizawa:2016jji}, as 
could be the case for example if they undergo dynamical interactions in dense stellar clusters \cite{OLeary:2008myb,Samsing:2018isx} or experience strong interactions with an AGN accretion disk \cite{Bellovary:2015ifg,Bartos:2016dgn,Stone:2016wzz}.
On the other hand, stellar-mass BBHs are not expected to present large eccentricities if they form via isolated binaries \cite{Belczynski:2016obo}.
The possible detection of large eccentricities has been in fact proposed as a way to distinguish BBH formation channels \cite{Breivik:2016ddj,Rodriguez:2016vmx}.
A LISA joint detection of systems with observable peculiar accelerations and large eccentricities would constitute a strong evidence for BHBs forming in AGN disks or globular clusters.
Note that as long as the peculiar acceleration is assumed to be constant over the duration of the LISA mission and sufficiently weak to not influence the BBH internal dynamics, the results presented here are expected to hold even in the presence of non-negligible eccentricity, as the motion of the center of mass of the BBH system can be decoupled from the intrinsic evolution.

Low mass binaries with a relatively short time to coalescence (we put a cutoff of 10 years from the start of LISA observation) can also be detected by Earth-based GW detectors at merger. We therefore also investigated how the acceleration effect may be measured in the case of joint LISA/Earth-based detection. We find that the number of events for which the effect becomes observable increases, reaching about 40\% of all detected events in the highest acceleration case.
As expected, the average uncertainty on the peculiar acceleration also improves for these joint observations, suggesting that the measurement of peculiar accelerations may constitute an interesting science case for multiband searches between LISA and Earth-based interferometers such as LIGO and Virgo.

For our analyses we assumed a LISA noise sensitivity curve as given in \cite{Audley:2017drz}.
The high frequency part of this curve is based on a single link optical measurement system (OMS) noise of 10 pm$/\sqrt{\rm Hz}$.
Note that in the LISA Science Requirements Document \cite{SRD} a margin has been inserted on this noise contribution, increasing it up to 15 pm$/\sqrt{\rm Hz}$.
Since our results are based on LISA observations at the high end frequencies of its sensitivity curve, the possible loss of sensitivity induced by this margin would affect them.
As a rough estimate we are expecting the sensitivity to worsen by a factor of 1.5, and thus SNRs and
measurement errors to degrade by the same factor. This would entail the loss of about 30\% (50\%) of the stellar-mass BHBs detections,
and about 55\% (22\%) of the events with acceleration measurements with error less than 100\%, for LISA-only (coincident) detections.
Nevertheless, the number of events with acceleration measured to better than 10\% in coincident LISA-ground based detections remains unchanged (such events are characterised by high SNR, thus the loss in sensitivity is less severe on their measurement).

We also studied whether the waveform parameters would be biased if the acceleration effect is ignored in the data analysis procedure. We computed a simple estimate of the bias on each waveform parameter, and studied how many events would experience a non-negligible bias. 
Our results clearly show that binaries possessing rather large peculiar accelerations (scenarios 3 to 5) will be biased even for a LISA mission duration of 4 years.
We also found that binaries with a time to coalescence similar to the LISA mission lifetime will have a higher probability of being biased.
The bias therefore follows the same parameter dependence of the measurement of the acceleration effect, but many events for which the peculiar acceleration would not be measurable are instead biased.
This suggests caution when performing data analysis in the parameter space of stellar-mass BHB systems coalescing close to the LISA lifetime, since for these systems, ignoring the peculiar acceleration could lead to wrong estimates of the BHB physical parameters.

Furthermore, we have found that the parameter most often biased is the luminosity distance of the system, which is a fundamental quantity for cosmological applications of GW observations, in particular standard siren analyses.
If the peculiar acceleration is not correctly taken into account, large systematics could affect the estimates of the cosmological parameters, e.g.~$H_0$, derived from LISA stellar-origin BHBs.
The time to coalescence can also be biased: this occurs for a smaller fraction of events than the luminosity distance, but for lower values of the peculiar acceleration, suggesting caution also when estimating the statistical error on this parameter without inserting the acceleration in the waveform.
If present, such a bias might affect multi-messenger and multiband searches which rely on an accurate estimation of the merger time of BHBs.

Finally, it is also interesting to note that the peculiar acceleration effect induces a perturbation on the phase of the gravitational waveform which has the same -4PN dependence in frequency of the effect given by a possible running of the gravitational constant \cite{Yunes:2009bv}.
In particular, if we compare the effect on the waveform phase given by $\dot{G}$ (namely the time variation of the gravitational constant $G$) as computed in \cite{Yunes:2009bv}, with the acceleration parameter $\alpha$ given in Eq.~\eqref{eq:alphadef}, we find
\begin{equation}
	\dot{G} = - \frac{2 \alpha \nu}{M_z} = -2 G Y_c \,.
\end{equation}
This result implies that studies of the running of the gravitational constant with GWs, and in particular with stellar mass BHBs, might be strongly affected by the possible presence of large peculiar accelerations. The investigation of these issues is left for future works.

\acknowledgments
This project has received funding (to E.~B.) from the European Research Council (ERC) under the European Union's Horizon 2020 research and innovation programme (grant agreement no.~GRAMS-815673; project title ``GRavity from Astrophysical to Microscopic Scales'').
This work was also supported by the H2020-MSCA-RISE-2015 Grant No.~StronGrHEP-690904.
C.~B.~acknowledges  support  by  the  Swiss  National  Science  Foundation.
This work was supported by the center National d'{\'E}tudes Spatiales, by the COST Action CA15117 ``Cosmology and Astrophysics Network for Theoretical Advances and Training Actions'' through a Short Term Scientific Mission (STSM), and by the COST Action CA16104 ``Gravitational waves, black holes and fundamental physics'' (GWverse). COST actions are supported by COST (European Cooperation in Science and Technology).


\appendix*

\section{Derivation of the acceleration effect in the gravitational waveform}

In this appendix we show in detail how the gravitational waveform produced by a binary inspiral is perturbed by the acceleration of the CoM of the binary system.
We reproduce the results of \cite{bonvin-2017} but correct a sign mistake in the wave phase and add a contribution in the wave amplitude.

\subsection{Derivation of the redshift perturbation}

\label{app:redshift}

We first derive the time dependence of the redshift for a GW source, expanding over suitable small quantities.
We start from the general definition of the redshift as measured by a cosmological observer in an FRW universe (see e.g.~\cite{peacock:1999})
\begin{equation}
	1+z = \frac{a(t_{\rm obs})}{a(t_{\rm src})} \left[ 1 + \frac{v_{\rm src}^\shortparallel(t_{\rm src})}{c} - \frac{v_{\rm obs}^\shortparallel(t_{\rm obs})}{c} \right] \,,
	\label{eq:001}
\end{equation}
where $v_{\rm src}^\shortparallel = \mathbf{n} \cdot \mathbf{v}_{\rm src}$ and $v_{\rm obs}^\shortparallel = \mathbf{n} \cdot \mathbf{v}_{\rm obs}$ are the peculiar velocities of the source and the observer (i.e.~the velocity that the objects have with respect to the Hubble flow), respectively, along the line of sight (identified with the unit vector $\mathbf{n}$), $a(t)$ is the scale factor and we explicitly write all dependences on the source local time $t_{\rm src}$ and the observer local time $t_{\rm obs}$.
Here we ignore the contributions from gravitational redshift and integrated Sachs-Wolfe since, as shown in~\cite{bonvin-2017}, they are sub-dominant. We consider $v_{\rm src}^\shortparallel/c \ll 1$ and $v_{\rm obs}^\shortparallel/c \ll 1$ as the non-relativistic astrophysical peculiar velocities of the source and the observer.
Following \cite{bonvin-2017} we want to expand the time dependent quantities in the r.h.s.~of Eq.~\eqref{eq:001}, which are assumed to be slowly varying in time, around a fixed chosen time in the source rest frame $t^*_{\rm src}$ and its corresponding time in the observer rest frame $t^*_{\rm obs}$.
In other words ${t}_{\rm obs} = t^*_{\rm obs} + ( {t}_{\rm obs} - t^*_{\rm obs} ) = t^*_{\rm obs} + \delta t_{\rm obs} $, where the time interval $\delta t_{\rm obs}$ is such that the functions in Eq.~\eqref{eq:001} vary slowly during this interval.
Similarly, we can expand $t_{\rm src}$ around $t^*_{\rm src}$, using $\delta t_{\rm src} = {t}_{\rm src} - t^*_{\rm src}$ as a small time interval.
In what follows, we will expand the time-dependent functions linearly in $\delta t_{\rm obs}$ and $\delta t_{\rm src}$, and keep only the terms at first order in $v_{\rm src}^\shortparallel/c$ and $v_{\rm obs}^\shortparallel/c$.

Given the assumptions above, Eq.~\eqref{eq:001} can be expanded as
\begin{widetext}
\begin{align}
1+z &\simeq \left[ \frac{a(t^*_{\rm obs}) + \dot{a}(t^*_{\rm obs}) ({t}_{\rm obs} - t^*_{\rm obs})}{a(t^*_{\rm src}) + \dot{a}(t^*_{\rm src}) ({t}_{\rm src} - t^*_{\rm src})} \right]
\left[ 1 + \frac{v_{\rm src}^\shortparallel(t^*_{\rm src})}{c} + \frac{\dot{v}_{\rm src}^\shortparallel(t^*_{\rm src})}{c} ({t}_{\rm src} - t^*_{\rm src}) - \frac{v_{\rm obs}^\shortparallel(t^*_{\rm obs})}{c} - \frac{\dot{v}_{\rm obs}^\shortparallel(t^*_{\rm obs})}{c} ({t}_{\rm obs} - t^*_{\rm obs}) \right] \nonumber\\
&\simeq \frac{a(t^*_{\rm obs})}{a(t^*_{\rm src})} \Big[1+ H(t^*_{\rm obs}) ({t}_{\rm obs} - t^*_{\rm obs}) - H(t^*_{\rm src}) ({t}_{\rm src} - t^*_{\rm src}) \Big] \left[ 1 + \frac{v_{\rm src}^\shortparallel(t^*_{\rm src})}{c} - \frac{v_{\rm obs}^\shortparallel(t^*_{\rm obs})}{c} \right] \nonumber\\
& \hspace{7cm} \times \left[ 1 + \frac{\dot{v}_{\rm src}^\shortparallel(t^*_{\rm src})}{c} ({t}_{\rm src} - t^*_{\rm src}) - \frac{\dot{v}_{\rm obs}^\shortparallel(t^*_{\rm obs})}{c} ({t}_{\rm obs} - t^*_{\rm obs}) \right] \,.\label{eq:app001}
\end{align}
\end{widetext}
At zeroth order in the time evolution 
this expression reads
\begin{equation}
1+z \simeq \frac{a(t^*_{\rm obs})}{a(t^*_{\rm src})} \left[ 1 + \frac{v_{\rm src}^\shortparallel(t^*_{\rm src})}{c} - \frac{v_{\rm obs}^\shortparallel(t^*_{\rm obs})}{c} \right] \equiv 1 + z_* \,,
\end{equation}
where we define $z_*=z(t^*_{\rm obs},t^*_{\rm src})$ as the redshift at the time $t^*$. Note that $z_*$ contains both the background expansion and the Doppler contribution.
We can use this relation at the lowest order to express $({t}_{\rm src} - t^*_{\rm src})$ in terms of $({t}_{\rm obs} -  t^*_{\rm obs})$:
\begin{align}
{t}_{\rm src} -  t^*_{\rm src} &\simeq \frac{{t}_{\rm obs} -  t^*_{\rm obs}}{1 + z_*} \,.
\end{align}
Inserting this into Eq.~\eqref{eq:app001} we find
\begin{equation}
1+z =  (1 + z_*) \Big[1+ 2 Y(z_*) ({t}_{\rm obs} -  t^*_{\rm obs}) \Big] \,,
\end{equation}
where we have defined
\begin{align}
Y(z_*) &\equiv  \frac{1}{2} \left[H(t^*_{\rm obs}) - \frac{H(t^*_{\rm src})}{1+z_*} + \frac{\dot{v}_{\rm src}^\shortparallel(t^*_{\rm src})}{c(1+z_*)} - \frac{\dot{v}_{\rm obs}^\shortparallel(t^*_{\rm obs})}{c}\right] \,.
\end{align}
We want to rewrite ${t}_{\rm obs} - t^*_{\rm obs}$ in terms of the time \textit{to} coalescence $\tau_{\rm obs} = t_c - t_{\rm obs}$, where $t_c$ is the time \textit{at} coalescence.
By noticing that
\begin{equation}
{t}_{\rm obs} - t^*_{\rm obs} = {t}_{\rm obs} - t_c + t_c - t^*_{\rm obs} = -\tau_{\rm obs} + \tau^*_{\rm obs} \,,
\end{equation}
we obtain
\begin{equation}
1+z = (1 + z_*) \Big[1- 2 Y(z_*) (\tau_{\rm obs} - \tau^*_{\rm obs}) \Big] \,.
\label{eq:zstar}
\end{equation}
For binaries that are observed close to the coalescence time $t_c$, a convenient choice is to take $t^*_{\rm obs}=t_c$ which yields $\tau^*_{\rm obs}=0$ and consequently Eq.~\eqref{zc}.

Note that another possibility would be to choose $t^*_{\rm obs}$ as the time when LISA starts taking data $t^*_{\rm obs}=t_{\rm st}$. For the binary systems analyzed in this work, $\tau_{\rm st}=t_c-t_{\rm st}$ is at most of the order of thousands of years, so that $\tau_{\rm st}Y(z_{\rm st})\ll 1$.
At first order we can therefore rewrite eq.~\eqref{eq:zstar} as
\begin{equation}
1+z = (1 + z_{\rm st}) \Big[ 1 + 2 Y_{\rm st} \tau_{\rm st} \Big] \Big[1- 2 Y_{\rm st} \tau_{\rm obs} \Big] \,.
\end{equation}
Since $\tau_{\rm st}Y_{\rm st}$ is constant and much smaller than one, we can reabsorb it into $z_{\rm st}$, i.e.~define an effective redshift 
\begin{equation}
1+\tilde z_{\rm st}\equiv(1+z_{\rm st})\Big[1 + 2 Y_{\rm st} \tau_{\rm st} \Big]\,,
\end{equation}
and we obtain
\begin{equation}
1+z = (1 + \tilde z_{\rm st}) \Big[1- 2 Y_{\rm st}\tau_{\rm obs}\Big] \,.
\label{eq:zstarbis}
\end{equation}
Comparing this with eq.~\eqref{zc} we see that the expressions are equivalent at lowest order in $Y\tau_{\rm obs}$. Indeed the constant redshift $z_c$ and $\tilde z_{\rm st}$ are not observable, since they are degenerated with the  mass, and therefore the difference between them does not matter. Furthermore the difference between $Y_c$ and $Y_{\rm st}$ is  of second order in the expansion, since it is due to the evolution of $Y$ between the time when we start observing and the coalescence time. Therefore, for the binaries we are interested in, one can equivalently set $t_{\rm obs}^*=t_{\rm st}$ or $t_{\rm obs}^*=t_{c}$.

\subsection{Derivation of the perturbed phase}

\label{app:phase}

Now that we know how the redshift changes over the time of observation of the binary, we want to understand how this affects the gravitational waveform as measured by the observer.
We start by deriving first the effect in the phase and then we will consider how the amplitude becomes perturbed.

The GW frequency evolution equation at the source is given by \cite{Blanchet-LR,Maggiore:1900zz}
\begin{equation}
\frac{d f_{\rm src}(t_{\rm src})}{d t_{\rm src}} = \frac{96}{5} \pi^{8/3} \left( \frac{G \mathcal{M}}{c^3} \right)^{5/3} f_{\rm src}(t_{\rm src})^{11/3} \,,
\label{eq:f_ev_src}
\end{equation}
where $f_{\rm src}$ is the GW frequency emitted by the source and $\mathcal{M}$ is the chirp mass.
At the observer both frequencies and time intervals are redshifted
\begin{equation}
f_{\rm src} = [1 + z(t_{\rm obs})] f_{\rm obs} \quad\mbox{and}\quad dt_{\rm src} = \frac{dt_{\rm obs}}{1+z(t_{\rm obs})} \,,
\label{eqapp:f_und_dt}
\end{equation}
where $z(t_{\rm obs})$ is given by Eq.~\eqref{eq:zstar}.
Substituting these expressions into Eq.~\eqref{eq:f_ev_src} and defining the new function
\begin{equation}
g(t_{\rm obs}) = [1 + z(t_{\rm obs})] f_{\rm obs} \,,
\end{equation}
one finds
\begin{equation}
\frac{d g(t_{\rm obs})}{d t_{\rm obs}} = \frac{96}{5} \pi^{8/3} \left( \frac{G \mathcal{M}}{c^3} \right)^{5/3} \frac{g(t_{\rm obs})^{11/3}}{1 + z(t_{\rm obs})}  \,.
\end{equation}
Integrating this expression between $t_{\rm obs}$ and $t_c$ (time of coalescence) and using that the frequency diverges at $t_c$, so that $g(t_c) \rightarrow \infty$, one obtains
\begin{align}
g(t_{\rm obs})^{-8/3}& = \frac{256}{5} \pi^{8/3} \left( \frac{G \mathcal{M}}{c^3} \right)^{5/3} \frac{1}{1 + z_*}\\
&\times \Big[ \tau_{\rm obs} + \Big((t_{\rm obs}-t^*_{\rm obs})^2-(t_c-t^*_{\rm obs})^2 \Big) Y_* \Big] \,.\nonumber
\end{align}
Note that this equation is valid only if the scale factor and the source peculiar velocity do not vary too much between $t_{\rm obs}$ and $t_c$. Choosing as before $t^*_{\rm obs}=t_c$, we obtain (we use here the notations of Eq.~\eqref{zc})
\begin{equation}
g(t_{\rm obs})^{-8/3} = \frac{256}{5} \pi^{8/3} \left( \frac{G \mathcal{M}}{c^3} \right)^{5/3} \frac{\tau_{\rm obs}}{1 + z_c} \left[ 1 + Y_c\, \tau_{\rm obs} \right] \,.
\end{equation}
Here and in what follows we keep only terms at first order in $Y_c\, \tau_{\rm obs} \ll 1$.
Rewriting everything in terms of $f_{\rm obs}$ yields
\begin{multline}
f_{\rm obs}(\tau_{\rm obs}) = \left(\frac{5}{256 \tau_{\rm obs}}\right)^{3/8} \frac{1}{\pi} \left(\frac{G \mathcal{M}_z}{c^3}\right)^{-5/8} \times \\
\times\left[ 1 + \frac{13}{8} Y_c\, \tau_{\rm obs} \right] \,,
\label{eq:f_tau}
\end{multline}
where now $\mathcal{M}_z = (1 + z_c) \mathcal{M}$ is the redshifted chirp mass at the time of coalescence.
The GW phase at the observer $\Phi\sub{obs} = 2 \phi\sub{obs}$, where $\phi\sub{obs}$ is the orbital phase, is then given by
\begin{align}
	\Phi_{\rm obs}(t_{\rm obs}) &= \Phi_c + 2 \pi \int_{t_c}^{t_{\rm obs}} dt_{\rm obs}' f_{\rm obs}(t_{\rm obs}')\label{eq:Phi_tau}\\
		&= \Phi_c -2 \left( \frac{\tau_{\rm obs} c^3}{5 G \mathcal{M}_z} \right)^{5/8} \left[ 1 + \frac{5}{8} Y_c\, \tau_{\rm obs} \right] \,,\nonumber
	\end{align}
where $\Phi_c$ is the value of the phase at coalescence.

In the Fourier space the phase is given by (see e.g.~Eq.~(4.367) in \cite{Maggiore:1900zz})
\begin{equation}
	\Psi(f_{\rm obs}) = 2 \pi f_{\rm obs} \left[ t_c - \tau_{\rm obs}(f_{\rm obs})\right] - \Phi_{\rm obs}(f_{\rm obs}) -\frac{\pi}{4}  \,,
	\label{eq:Psi_def}
\end{equation}
where $\tau_{\rm obs}$ and $\Phi_{\rm obs}$ must be considered as functions of $f_{\rm obs}$, which in turn can be found from Eqs.~\eqref{eq:f_tau} and \eqref{eq:Phi_tau}.
Inverting Eq.~\eqref{eq:f_tau} one finds
\begin{multline}
	\tau_{\rm obs}(f_{\rm obs}) = \frac{5}{256} \pi^{-8/3} \left(\frac{G \mathcal{M}_z}{c^3}\right)^{-5/3} f_{\rm obs}^{-8/3} \\
	\times\left[ 1 + \frac{65}{768} \pi^{-8/3} \left(\frac{G \mathcal{M}_z}{c^3}\right)^{-5/3} Y_c\, f_{\rm obs}^{-8/3} \right]
	\,, 
	\label{eq:tau_f}
\end{multline}
which inserted into Eq.~\eqref{eq:Phi_tau} yields
\begin{multline}
	\Phi_{\rm obs}(f_{\rm obs}) = \Phi_c - \frac{1}{16} \left( \frac{\pi G \mathcal{M}_z}{c^3} \right)^{-5/3} f_{\rm obs}^{-5/3} \\
	\times\left[ 1 + \frac{25}{384} \pi^{-8/3} \left(\frac{G \mathcal{M}_z}{c^3}\right)^{-5/3} Y_c\, f^{-8/3} \right]  
	\,.
\end{multline}
Substituting these two results into Eq.~\eqref{eq:Psi_def} finally yields
\begin{multline}
	\Psi(f_{\rm obs}) = 2 \pi f_{\rm obs} t_c - \frac{\pi}{4} -\Phi_c + \frac{3}{128} \left( \frac{\pi G \mathcal{M}_z}{c^3} \right)^{-5/3} f_{\rm obs}^{-5/3}\\
	 + \frac{25}{32768} \left( \frac{ G \mathcal{M}_z}{c^3} \right)^{-10/3} \pi^{-13/3} Y_c\, f_{\rm obs}^{-13/3} \,.
	 \label{eq:app_pert_phase}
\end{multline}
This equation coincides with Eq.~\eqref{eq:pert_phase_FS} reported in the main body of the paper (for $n=2$).
To compare with Eq.~(44) of \cite{bonvin-2017}, we need to account for the fact that the redshifted chirp mass there (denoted by $\mathcal{M}_c$) is evaluated at the observer time $t_{\rm obs}$: $\mathcal{M}_c(t\sub{obs})\equiv \mathcal{M}_{z_{\rm obs}}=(1+z(\tau_{\rm obs})) \mathcal{M}$. This introduces an additional time variation into the phase. Rewriting $\mathcal{M}_{z_{\rm obs}}=(1+z_c)(1-2Y_c\tau_{\rm obs}) \mathcal{M}=(1-2Y_c\tau_{\rm obs})\mathcal{M}_{z_c}$, Eq.~(44) of~\cite{bonvin-2017} reduces to Eq.~\eqref{eq:app_pert_phase}.

\subsection{Derivation of the perturbed amplitude}

\label{app:amplitude}

The amplitude is defined as the part of the waveform which does not depend on geometrical factors and is thus common to both GW polarizations.  It is given by (see e.g.~Eq.~(4.29) of \cite{Maggiore:1900zz})
\begin{equation}
\mathcal{A}_c(f_{\rm src}) = \frac{4}{d_C(t_{\rm src}) c^4} \left( G \mathcal{M} \right)^{5/3} \left( \pi f_{\rm src} \right)^{2/3} \,,
\end{equation}
where $d_C$ denotes the comoving distance to the binary. Using Eq.~\eqref{frequency}, the amplitude can be written as a function of $f_{\rm obs}$
\begin{equation}
\label{Af0}
\mathcal{A}_c(f_{\rm obs}) = \frac{4}{d_L(t_{\rm src}) c^4} \left( G \mathcal{M}(1+z) \right)^{5/3} \left( \pi f_{\rm obs} \right)^{2/3} \,,
\end{equation}
where $d_L=(1+z)d_C$ is the luminosity distance of the binary.

Under the stationary phase approximation the common amplitude in the Fourier space reads (see e.g.~Eq.~(4.366) of \cite{Maggiore:1900zz})
\begin{equation}
	\tilde{\mathcal{A}}(f_{\rm obs}) = \frac{1}{2} \mathcal{A}_c(f_{\rm obs}) \left( \frac{2 \pi}{\ddot\Phi(\tau_{\rm obs}(f_{\rm obs}))} \right)^{1/2} \,,
	\label{eq:007}
\end{equation}
where
\begin{equation}
	\ddot\Phi(\tau_{\rm obs}) = \frac{d^2}{d \tau_{\rm obs}^2} \Phi(\tau_{\rm src}) = \frac{d^2}{d t_{\rm obs}^2} \Phi(t_{\rm src}) \,,
\end{equation}
since $d \tau_{\rm obs} = - dt_{\rm obs}$.
Using Eq.~\eqref{eq:Phi_tau} one finds
\begin{equation}
	\ddot\Phi(\tau_{\rm obs}) = \frac{3}{32} 5^{3/8} \left( \frac{G \mathcal{M}_z}{c^3} \right)^{-5/8} \tau_{\rm obs}^{-11/8} \left[ 1 - \frac{65}{24} Y_c\, \tau_{\rm obs} \right]
\end{equation}
and at first order in $Y_c\,\tau_{\rm obs}$ we thus obtain
\begin{multline}
	\left( \frac{2 \pi}{\ddot\Phi(\tau_{\rm obs})} \right)^{1/2} =\frac{8\sqrt{\pi}}{\sqrt{3}} 5^{-3/16}  \left( \frac{G \mathcal{M}_z}{c^3} \right)^{5/16} \tau_{\rm obs}^{11/16} \\
	\times\left[ 1 + \frac{65}{48} Y_c\, \tau_{\rm obs} \right] \,.
\end{multline}
Using Eq.~\eqref{eq:tau_f} we can rewrite this in term of the frequency as
\begin{multline}
\left( \frac{2 \pi}{\ddot\Phi(\tau_{\rm obs}(f_{\rm obs}))} \right)^{1/2} = \frac{1}{4} \sqrt{\frac{5}{6}} \pi^{-4/3} \left( \frac{G \mathcal{M}_z}{c^3} \right)^{-5/6} f_{\rm obs}^{-11/6} \\
\times\left[ 1 + \frac{65}{768} \left( \frac{G \mathcal{M}_z}{c^3} \right)^{-5/3} \pi^{-8/3} Y_c\, f_{\rm obs}^{-8/3} \right] \,. 
\label{eq:006}
\end{multline}

To this, we need to add the contribution from the luminosity distance in Eq.~\eqref{Af0}. As shown in~\cite{Bonvin:2005ps}, the luminosity distance is affected by matter perturbations. At high redshift the dominant contribution is the one from gravitational lensing, whereas at low redshift it is due to the source peculiar velocity
\begin{align}
d_L(z)=&(1+z(t_{\rm src}))\chi(t_{\rm src})\Bigg[1+\label{eq:dL}\\
&+\left(1-\frac{1}{\chi(t_{\rm src})\mathcal{H}(t_{\rm src})} \right)v_{\rm src}^\shortparallel(t_{\rm src})\nonumber\\
&-\int_0^{\chi(t_{\rm src})}d\chi\frac{(\chi(t_{\rm src})-\chi)}{2\chi\chi(t_{\rm src})}\Delta_\Omega(\phi+\psi) \Bigg]\, ,\nonumber
\end{align}
where $\chi(t_{\rm src})=c\int_0^{t_{\rm src}}dt/a$ denotes the background comoving distance to the source, $\Delta_\Omega$ is the transverse Laplacian and $\phi$ and $\psi$ are the two metric potentials. As before, we expand quantities at $t_{\rm src}$ around the time at coalescence $t_c$. We neglect in the following the contributions coming from the background evolution, since they are strongly subdominant with respect to the source peculiar acceleration. We neglect also the contribution due to the evolution of gravitational lensing. This contribution would indeed be proportional to the time variation of the metric potentials, $\dot{\psi}$ and $\dot{\phi}$. In a matter dominated universe, the potentials are constant in time, and these terms are exactly zero. The presence of a cosmological constant does induce a time variation of the potentials, but this time variation is much smaller than the time variation of the peculiar velocity and can therefore be safely neglected. We obtain then
\begin{align}
&d_L(z)=(1+z_c)(1-2Y_c\tau_{\rm obs})\chi_c\\
&\times\Bigg\{1+\left(1-\frac{1}{\HH_c\chi_c} \right) \big[v^\shortparallel_s(t_c)+\dot{v}^\shortparallel_s(t_c)(t_s-t_{c})\big]\nonumber\\
&-\int_0^{\chi(t_{\rm src})}d\chi\frac{(\chi(t_{\rm src})-\chi)}{2\chi\chi(t_{\rm src})}\Delta_\Omega(\phi+\psi) \Bigg\}\nonumber\\
&=d_L(z_c)\left\{1-2\left(2-\frac{1}{\HH_c\chi_c}\right)Y_c\tau_{\rm obs}\right\}\, ,\nonumber
\end{align}
where in the last line we have neglected terms that are second order in the perturbations.
Including this expression into~\eqref{Af0} we obtain
\begin{align}
&\mathcal{A}_c(f_{\rm obs}) = \frac{4}{d_L(z_c) c^4} \left( G \mathcal{M}_z \right)^{5/3}(\pi f_{\rm obs})^{2/3}\\
&\times \left[ 1 +2 \left(\frac{1}{3}  - \frac{1}{\HH_c\chi_c}\right) Y_c\, \tau_{\rm obs} \right] \,,\nonumber
\end{align}
where $\mathcal{M}_z=(1+z_c)\mathcal{M}$.
Using Eq.~\eqref{eq:tau_f} and expanding in $Y_c\, \tau_{\rm obs}$ we obtain
\begin{align}
&\mathcal{A}_c(f_{\rm obs}) = \frac{4}{d_L(z_c) c^4} \left(G \mathcal{M}_z\right)^{5/3}(\pi f_{\rm obs})^{2/3} \label{eq:005} \\
&\times\Bigg[ 1 +\frac{5}{384} \left( \frac{G \mathcal{M}_z}{c^3} \right)^{5/3}
\left(1-\frac{3}{\HH_c\chi_c}\right) Y_c\, (\pi f_{\rm obs})^{-8/3} \Bigg] \,.\nonumber
\end{align}

Finally substituting Eqs.~\eqref{eq:006} and \eqref{eq:005} into Eq.~\eqref{eq:007} one finds
\begin{align}
&\tilde{\mathcal{A}}(f_{\rm obs}) = \frac{1}{d_L(z_c)} \sqrt{\frac{5}{24 \pi^{4/3} c^3}} \left( G \mathcal{M}_{z} \right)^{5/6} f_{\rm obs}^{-7/6}\\
&\Bigg[ 1 + \frac{5\big(G \mathcal{M}_{z}/c^3\big)^{-5/3}\,Y_c}{128\big(\pi f_{\rm obs}\big)^{8/3}}\left(\frac{5}{2}-\frac{1}{\HH_c\chi_c} \right) \Bigg] \,,\nonumber
\end{align}
which agrees with Eq.~\eqref{eq:amp_corrected_FS} reported in the main body of the paper (recall that $\mathcal{M}_z=\nu^{3/5}M_z$).
Note that the acceleration correction in the amplitude computed here differs from the one reported in \cite{bonvin-2017} which didn't account for the velocity contribution in the luminosity distance.

In all the analyses of this paper we took into account this amplitude perturbation by computing $\HH_c\chi_c$ assuming a standard $\Lambda$CDM cosmology.

\bibliography{accel-paper}

\end{document}